\documentclass[letterpaper,11pt]{article}
\title{Liquid Welfare and Revenue Monotonicity \\ in Adaptive Clinching Auctions}

\author{Ryosuke Sato \thanks{Keio University, Yokohama, Japan. ryosuke\_sato@math.keio.ac.jp}}

\usepackage[top=1.2truein,bottom=1.2truein,left=1truein,right=1truein]{geometry}
\usepackage{algorithm}
\usepackage{algorithmic}
\usepackage{enumitem}
\usepackage{amsmath}
\usepackage{amsthm}
\usepackage{amsfonts}
\usepackage{booktabs}
\usepackage{multirow}
\usepackage[authoryear]{natbib}
\begin{document}
\theoremstyle{definition}
\newtheorem{definition}{Definition}[section]
\newtheorem{proposition}[definition]{Proposition}
\newtheorem{lemma}[definition]{Lemma}
\newtheorem*{main}{Main Theorem}
\newtheorem*{theorem2}{Theorem}
\newtheorem{theorem}[definition]{Theorem}
\newtheorem{corollary}[definition]{Corollary}
\newtheorem{remark}[definition]{Remark}
\newtheorem*{rem}{Remark}
\newtheorem{fact}[definition]{Fact}
\newtheorem*{claim}{Claim}
\newtheorem{observation}[definition]{Observation}
\newtheorem{assmp}[definition]{Assumption}
\newtheorem{example}[definition]{Example}

\maketitle
\begin{abstract}
This study explores the monotonicity of adaptive clinching auctions -- a key mechanism in budget-constrained auctions -- with respect to fluctuations in the number of bidders. Specifically, we investigate how the addition of new bidders affect efficiency and revenue. In a symmetric setting, where all bidders have equal budgets, we show that while the allocated goods and payments for many bidders decrease, overall both liquid welfare and revenue weakly increase. Our analysis also extends to scenarios where bidders arrive online during the auction. In contrast, for asymmetric budgets, we provide counterexamples showing that these monotonicity properties no longer hold. These findings contribute to a better theoretical understanding of budget-constrained auctions and offer insights into the behavior of adaptive clinching auctions in social networks, where new bidders emerge through information diffusion.
\end{abstract}

\section{Introduction}
\textit{Budget-constrained auctions} are a fundamental setting in auction theory, particularly for real-world applications such as ad auctions. The \textit{adaptive clinching auction} proposed by \citet{DLN2012} has played an important role in this context. Built on Ausubel's clinching framework \citep{A2004}, it is still the only budget-feasible mechanism that satisfies \textit{incentive compatibility}\ (IC), \textit{individual rationality} (IR), and \textit{Pareto optimality} (PO). Consequently, clinching auctions are now widely accepted as standard mechanisms in auctions with budgets and have inspired extensive additional research.

Recently, auctions with \textit{information diffusion} \citep{LHZZ2017} have gained significant attention, as transactions through \textit{social networks} have become increasingly common. These auctions are characterized by incentive design that encourages participants to spread information. Although the participation of new bidders is expected to improve  outcomes, it also increases competition. Following this research trend, \citet{XSK2022} introduced budget-constrained auctions in social networks and proposed an ascending auction that builds on the adaptive clinching auction. In their framework, when a bidder drops out, their neighboring potential bidders are invited to participate, which can be viewed as the {\it online arrival} of new bidders. The incentive design for information diffusion is effective in their mechanism, satisfying IC, IR, and non-wastefulness -- an efficiency concept introduced by~\citet{KBTTY2020}.

While non-wastefulness is valuable, it is relatively weak, thereby allowing for further improvement in the theoretical guarantees of the mechanism. Ideally, as more bidders engage through information diffusion, both efficiency and revenue should increase, ensuring that the spread of information leads to better outcomes. 
To assess whether efficiency and revenue remain monotonic with the arrival of new bidders, the effect of participation on clinching auctions must be analyzed.
This question is not only practically and fundamentally important, but it also extends previous research on the monotonicity of clinching auctions by \citet{BCMX2010} and \citet{GMP2020}. In this study, we investigate the monotonicity of adaptive clinching auctions in the presence of new~bidders.

\subsection{Our Contributions}
This study focuses on budget-constrained auctions of a single unit of an infinitely divisible good, a common setting for investigating monotonicity, as explored in prior studies \citep{BCMX2010,GMP2020}. We begin with the symmetric case, where all bidders have equal budgets. Such symmetric settings are economically well-motivated and exhibit stronger properties in clinching auctions \citep{DHH2013, FH2018}, thereby making them an appropriate starting point for our analysis. 

First, we show that adding a new bidder from the beginning causes the following~changes:
\begin{itemize}[left=5pt]
\item[(i)] The outcome is weakly decreasing for all bidders except one,
\item[(ii)] Both liquid welfare (LW) and revenue are weakly increasing, but social welfare (SW) may~not.
\end{itemize}
Property (i) captures the monotonicity of each bidder's outcome, while property (ii) reflects the monotonicity of key auction outcomes, such as LW and revenue. Here, LW  is an efficiency objective that extends SW by incorporating budgets \citep{DL2014,ST2013}. Further, property (ii) supports the extension of clinching auctions to social networks because it 
ensures better outcomes when new bidders participate through information diffusion. By repeatedly applying this result, this the monotonicity can be extended to the addition of multiple bidders.

In the proof, we formulate an explicit formula for the allocations and remaining budgets at any price using the {\it clinching interval}~\citep{DL2014}, which is the price interval where the good is actually traded. We then detect changes in the clinching interval and use the formula to compare outcomes before and after the input change. Independently, this technique may contribute to the research of \citet{DHH2013} and \citet{FH2018}, who investigated the efficiency under symmetric~bidders.

Building on these findings, our next focus is to determine how broadly the monotonicity of LW and revenue holds across various scenarios in symmetric settings. Consider the case where new bidders arrive online during the auction. In this situation, the remaining budgets of bidders may not always be equal. Consequently, the previous formula 
cannot be applied directly. 
By extending the formula to incorporate the online arrival of bidders in a key case, we show that both LW and revenue monotonicity remain valid. This suggests that information diffusion improves efficiency and revenue in the clinching auction with symmetric bidders.

Further, we investigate the asymmetric setting, where bidders have distinct budgets. The monotonicity of clinching auctions in such settings has been investigated by \citet{BCMX2010} and \citet{GMP2020}, as summarized in Table~\ref{comparison}. While their analyses focused on changes in a single parameter, ours must 
consider the addition of two parameters: the valuation and the budget of the new bidder. The interaction of these parameters influences the outcome in various ways. Notably, we find counterexamples showing that neither each bidder's outcome nor LW is monotonic. As for revenue monotonicity, the current evidence suggests that it remains an open question. Thus, we provide a counterexample for the {\it indivisible} case, which is more relevant for applications. 

\begin{table}[t]
\centering
\caption{Summary of studies on the monotonicity of adaptive clinching auctions}
\label{comparison}
\begin{tabular}{lcc}
\toprule
& Change in the Input Parameter& Monotonicity Focus  \\ 
\midrule
\citet{BCMX2010} &  Decrease Budget & Utility \\ 
\citet{GMP2020}& Increase Supply & Allocation \\  
This Study& Add Bidder & LW and Revenue \\ 
\bottomrule
\end{tabular}
\end{table}

\subsection{Related Works}
In budget-constrained auctions, 
it is well known that no mechanism can achieve a constant approximation to the optimal SW while satisfying IC and IR. 
\citet{DLN2012} pioneered the use of PO as an alternative efficiency goal. 
\citet{GL2019} characterized the the space of mechanisms that satisfies these desirable properties 
in the market with two bidders and heterogeneous goods. 
LW \citep{DL2014,ST2013} is now widely recognized as a standard efficiency objective for budget-constrained auctions (e.g., \citep{FT2023, FLP2019,LX2015,LX2017}).
For clinching auctions, \citet{DL2014} showed that the adaptive clinching auction achieves a 1/2-approximation to the optimal LW and recent studies \citep{S2023,HS2025} have extended this result to polymatroidal settings. Furthermore,  \citet{DHH2013} and \citet{FH2018} presented stronger efficiency guarantees for clinching auctions in symmetric settings. The monotonicity of the LW was not addressed in these studies.

\citet{LHZZ2017} initiated the study of auctions in social networks, proposing a mechanism that achieves higher revenue than the VCG mechanism~\citep{V1961,C1971,G1973} extended to this setting. For efficiency guarantees, \citet{LHGZ2022} showed that any efficient mechanism satisfying IC and IR results in a seller's deficit, necessitating a weaker notion of efficiency. Although these studies focused on the case of a single unit, the frameworks have been applied to more general settings, including multi-unit auctions~\citep{ZLXHJ2018,TKTY2019,KBTTY2020,LLZ2023}, budget-constrained auctions~\citep{XSK2022}, and double auctions~\citep{LCZ2024,XHZ2019}. For a more detailed overview of this research trend, see \citet{Z2022}. Also, 
the arrival of new bidders has been a central topic in online and dynamic market settings, 
see \citet{P2007} for more details.

Revenue monotonicity (e.g., \citep{M2004,RCL2011}) is a well-studied property.  
It ensures that revenue increases monotonically in response to changes in inputs or other factors.  
Among various notions, bidder revenue monotonicity is the most relevant for auctions in social networks,  
as \citet{KBTTY2020} considered this property in this context.  
For the VCG mechanism, \citet{AM2002} and \citet{DRS2012} have characterized  
the conditions under which this property holds.  
To the best of our knowledge, this property has not been considered for clinching auctions.

\paragraph{Organization of this paper.}
The remainder of the paper is organized as follows.  
Section~2 presents our model.  
Section~3 introduces the adaptive clinching auction and provides some useful properties under the symmetric setting.  
Section~4 investigates monotonicity under the addition of a new bidder.  
Section~5 extends the results of Section~4 to a setting where bidders arrive online.  
Section~6 discusses the challenges of the asymmetric setting.  
Section~7 provides some concluding remarks.  
All omitted proofs are given in the appendix.

\paragraph{Notation.}
Let $\mathbf R_+$ denote the set of non-negative real numbers. 
We often denote a singleton $\{i\}$ by $i$ and a set $\{1,2,\ldots,k\}$ by $[k]$. For a vector $x\in \mathbf R^N_+$, we often denote $x(i)$ by $x_i$, and write as $x= (x_i)_{i\in N}$. We also define $x_{-i}$ by $x_{-i}:=(x_j)_{j\in N\setminus i}$.

\section{Our Model}
Consider a market with $n$ bidders $(n \geq 2)$ and a seller selling a single unit of a divisible good. 
We often denote a singleton $\{i\}$ by $i$ and a set $\{1,2,\ldots,k\}$ by $[k]$. 
Each bidder $i$ has a valuation of $v_i \in \mathbf{R}_+$ for one unit of the good and strategically reports a bid $v'_i \in \mathbf{R}_+$. 
Each bidder also has a public budget of $B_i\in \mathbf{R}_+$, which represents the maximum possible payment.
Define $N := [n]$, $v := \{v_i\}_{i \in N}$, $v' := \{v'_i\}_{i \in N}$, and $B := \{B_i\}_{i \in N}$.
Bidders are assumed to have different valuations and are in descending order of their valuations,
i.e., $v_1\geq v_2\geq \ldots\geq v_n$.

An {\it outcome} $(x, \pi)$ consists of  an {\it allocation} $x := (x_i)_{i \in N}$ and a {\it payment} $\pi := (\pi_i)_{i \in N}$, where $x_i \in \mathbf{R}_+$ denotes the amount of good allocated to bidder $i$ and $\pi_i \in \mathbf{R}_+$ denotes the payment of $i$. The outcome must satisfy the following conditions: $\sum_{i \in N} x_i = 1$ and $\pi_i \leq B_i$ for each $i \in N$. A  {\it mechanism} $\mathcal M$ is a map that determines the outcome based on the bids.
The  {\it utility} $u_i$ of bidder $i$ is given by $u_i(v', \mathcal{M}) := v_i x_i - \pi_i$ if $\pi_i \leq B_i$ and $-\infty$ otherwise, where $(x, \pi)$ is the outcome from the mechanism $\mathcal M$ under $v'$.

We consider a budget-feasible mechanism $\mathcal M$ that satisfies the following desirable properties:
\begin{itemize}[left=5pt]
\item \textbf{Incentive Compatibility (IC):} It is the best strategy for each bidder to report their true valuation, i.e., $u_i((v_i, v'_{-i}), \mathcal{M}) \geq u_i(v', \mathcal{M})$ for each $i$ and $v'$.
\item \textbf{Individual Rationality (IR):} There is a bid such that each bidder receives non-negative utility. If IC holds, IR is expressed as $u_i((v_i, v'_{-i}), \mathcal{M}) \geq 0$ for each $i$ and $v'$.
\end{itemize}
These properties ensure truthful bidding and voluntary participation of bidders, respectively. Given an outcome $(x, \pi)$, the objectives focused in this paper are described as follows:
\begin{itemize}[left=5pt]
\item \textbf{Social Welfare (SW):} A standard efficiency objective, defined by ${\rm SW}(x):=\sum_{i \in N} v_i x_i$. It  represents the total sum of willingness-to-pay (i.e., valuations) of bidders for their allocated~good.
\item \textbf{Liquid Welfare (LW):} A natural extension of SW that incorporates the ability-to-pay (i.e., budget constraints), defined by ${\rm LW}(x) := \sum_{i \in N} \min(v_i x_i, B_i)$. 
It represents the total sum of admissibility-to-pay of bidders for their allocated good.
\item \textbf{Revenue:} The seller's revenue, denoted by~${\rm REV}(\pi):= \sum_{i \in N} \pi_i$.
\end{itemize}

\section{Adaptive Clinching Auctions}
In this section, we introduce the continuous form of the adaptive clinching auction by \citet{BCMX2010} and provide its fundamental properties. Furthermore, we explore certain useful properties that hold in the symmetric setting. Note that this mechanism is known to satisfy IC and thus we assume in the rest of this paper that all bidders bid truthfully, i.e., $v'_i=v_i$ for each $i\in N$.

\subsection{Overview of the Mechanism}
The adaptive clinching auction \citep{BCMX2010,DLN2012} is an ascending auction inspired by Ausubel's clinching framework \citep{A2004}. Intuitively, in this mechanism, the price clock gradually increases from zero, and at each price, each bidder wins (or \textit{clinches}) the good at the current price if the total demand of other bidders is less than the amount of remaining good. 

 The mechanism is described in Algorithm \ref{alg:algorithm}, using the following notation:
\begin{algorithm}[tb]
    \caption{Adaptive Clinching Auction:\ ${\rm Adp}(N, v,B)$}
    \label{alg:algorithm}
    \begin{flushleft}
     \noindent \textbf{Input}: $N, v, B$ (Ordered so that $v_i> v_j$ for each $i< j$.)\\
    \noindent\textbf{Output}: $x^{\rm f}:=(x^{\rm f}_i)_{i\in N}$ and $\pi^{\rm f}:=(\pi^{\rm f}_i)_{i\in N}$.
    \end{flushleft}
    \begin{algorithmic}[1] 
        \STATE $x_i(0)=0, b_i(0):=B_i\, (i\in N), A(0):=N, {\rm and}\, p:=0$.
        \FOR{$k=n,n-1,\ldots,1$}
        \STATE Continuously increase $p$ while $p<v_k$. 
        Along the way, $x_i(p)$ and $b_i(p)$\ $(i\in N)$ are changed~as: 
        \begin{itemize}
	\item For $i\in C(p)$, $\partial_{p} x_i(p):= S(p)/p$, $\partial_{p}b_i(p):=-S(p)$. 
        \item For $i\notin C(p)$, $\partial_{p} x_i(p)=\partial_{p} b_i(p):=0$.
	\end{itemize}
	\STATE $A(v_k):=[k-1]$.
	\FOR{$i\in A(v_k)$}
	\STATE $\delta_i:= \max(0, S(v_k-)-  \sum_{j\in A(v_k)\setminus i}b_j(v_k-)/v_k )$. 
	\STATE $x_i(v_k):=x_i(v_k-)+\delta_i$,\ \ $b_i(v_k):=b_i(v_k-)-v_k\delta_i$.
	\ENDFOR
	\FOR{$i\notin A(v_k)$}
	\STATE $x_i(v_k):=x_i(v_k-)$,\ \ $b_i(v_k):=b_i(v_k-)$.
        \ENDFOR
        \ENDFOR
        \STATE  $x^{\rm f}_i:=x_i(v_1), \pi^{\rm f}_i:=B_i-b_i(v_1)$\ \ ($i\in N$).
    \end{algorithmic}
\end{algorithm}

\begin{itemize}[left=5pt]
\item $p$ is the \textit{price clock}, which starts at zero and increases during the auction, representing the transaction price per unit.
\item For each bidder $i$, let $x_i(p)$ denote the amount of their allocated good and $b_i(p)$ denote their remaining budget at price $p$ in the mechanism.
Initially, the former is set to zero and the latter to $B_i$. They are updated according to their transactions. They are right continuous at all points $p$, and left continuous at all points $p\notin \{v_1,v_2,\ldots, v_n\}$.
\item $A(p)$ is the set of \textit{active} bidders at price $p$, defined by $A(p) :=\{i \in N \mid p < v_i\}$. A bidder $i$ is said to be active if $p < v_i$ and \textit{dropped} otherwise. During line 3, the set $A(p)$ remains unchanged. 
\item $S(p):=1-\sum_{i\in N}x_i(p)$ is the amount of remaining good at price $p$.
\item The demand of bidder $i$ at price $p$ is defined as $b_i(p)/p$ if $i$ is active and $0$ otherwise. 
\item $C(p)$ is the \textit{clinching set} at price $p$, including bidders who clinch the good when the total demand of other bidders decreases, i.e., $C(p):=\{i\in A(p)\mid S(p)= \sum_{j\in A(p)\setminus i}b_j(p)/p \}$.
\item $\delta_i$ denotes the amount of good clinched by active bidder $i$ in line 7 
just after a bidder drops out of the auction in line 4.
\item For any function $f$, let $f(p-)$ denote the left-hand limit of $f$ at $p$ and $\partial_p f(p)$ denote the right-derivative of $f$ at $p$.
\item $(x^{\rm f}, \pi^{\rm f})$ denotes the final allocation of the mechanism.
\end{itemize}

In Algorithm \ref{alg:algorithm}, if the total demand of bidders except $i$ 
is less than the amount of remaining good, 
i.e., $\sum_{j\in A(p)\setminus i}b_j(p)/p<S(p)$, bidder $i$ clinches the amount of good equal to the difference. The transactions in lines 3 and 6 are conducted whenever the price is increased. 
The auction proceeds by keeping $S(p)\leq \sum_{j\in A(p)\setminus i}b_j(p)/p$ in lines 3 and 11 for each $i$ and $p$. Once this inequality holds in equality for some bidder $i$ and price $q$, then $i$ enters the clinching set $C(q)$ and the equality holds for $i$ at every price $p\geq q$. 
\citet{BCMX2010} showed that if $C(p)\neq \emptyset$, then it holds 
\begin{equation}
\label{clinchingset}
C(p)=\{i\in A(p)\mid b_i(p)=\max_{j\in A(p)}b_j(p)\}.
\end{equation}
This implies that the remaining budgets are equal for all bidders in the clinching set. 

Our analyses revolve around two key concepts: the {\it clinching interval} and the {\it critical bidder}. The clinching interval is the price range at which the good is traded, and the critical bidder is the one with~the highest valuation among those who have not exhausted their entire budgets.
\begin{definition}[\citet{DL2014}]
\label{interval}
The clinching interval is the interval $[p_s, p_f]$ of prices, where 
$p_s=\min\{q\mid \exists i\ \, {\rm s.t.}\ S(q)=\sum_{j\in A(q)\setminus i}b_j(q)/q\}\  {\rm and}\ 
p_f=\min\{p\mid \sum_{i\in N}x_i(p)=1\}.$
\end{definition} 
\begin{definition}
\label{critical}
The critical bidder refers to bidder $\kappa$, where $\kappa := \left| \{ i \mid \pi^{\rm f}_i = B_i \} \right| + 1$, i.e., the first bidder whose budget is not exhausted.\footnote{This concept was appeared in previous studies (e.g., \citep{BCMX2010,DL2014})
but not explicitly defined.}
\end{definition} 

This mechanism is known to satisfy several desirable properties. 
The following theorem summarizes key findings of previous studies \citep{BCMX2010,DLN2012,DL2014}.
\begin{theorem}
\label{properties_sec5}
In Algorithm~\ref{alg:algorithm}, the following holds:
\begin{itemize}
\item[{\rm (i)}] It satisfies all of IC, IR, and PO. 
Moreover, it achieves 1/2-approximation to the optimal LW.
\item[{\rm (ii)}] At the end of the auction, it holds $\sum_{i\in N}x^{\rm f}_i=1$, i.e., all the good is sold.
\item[{\rm (iii)}] It holds $\kappa=1$ if and only if $v_2< B_1$. 
In this case, it holds $p_s=p_f=v_2$, $(x^{\rm f}_1, \pi^{\rm f}_1)=(1, v_2)$, and $(x^{\rm f}_i,\pi^{\rm f}_i)=(0,0)$ for all $i>1$.  
If $\kappa>1$, it holds $p_s\leq p_f=v_\kappa$ and 
\begin{align*}
(x^{\rm f}_i, \pi^{\rm f}_i)=
\begin{cases}
(x_i(p_f-)+b_i(p_f-)/p_f, B_i) &{\rm if}\  i<\kappa,\\
(x_i(p_f-), B_i-b_i(p_f-)) &{\rm if}\  i=\kappa,\\
(0,0) &{\rm if}\ i>\kappa.
\end{cases}
\end{align*}
\end{itemize}
\end{theorem}
Note that property (iii) characterizes the final allocation, implying that all the remaining good is allocated at $p=v_{\kappa+1}$ if $\kappa=1$ and $p=v_{\kappa}$ otherwise. This price corresponds to $p_f$.

\subsection{Properties for Symmetric Settings}
An economically well-motivated and tractable case is the symmetric setting, where all bidders have equal budgets, i.e., $B_i = \beta$ for each $i$ and some $\beta > 0$. Now we show that Algorithm~\ref{alg:algorithm} exhibits a much simpler structure in this setting, which plays a crucial role in Sections 4 and 5.

Since the bids are only used to determine when bidders drop out,  
all active bidders are treated equally and thus their remaining budgets are equal, 
i.e., $b_i(p)=b_j(p)$ for each $i,j\in A(p)$.
Using this and (\ref{clinchingset}), we obtain the following observations: 
\begin{itemize}
\item $\sum_{j\in A(p)\setminus i}b_j(p)/p=(|A(p)|-1)b_i(p)/p$ 
for each $i\in A(p)$. 
\item $C(p)=A(p)$ if $p\geq p_s$ and $C(p)=\emptyset$ otherwise.
\end{itemize}
By these useful properties, clinching intervals can be precisely calculated.
\begin{lemma}
\label{p_s_p_f_common}
It holds $\max(v_{\kappa+1}, (\kappa-1) \beta)= p_s<\kappa \beta$.
\end{lemma}

Combining this lemma with Theorem \ref{properties_sec5} (iii), we have
\begin{align}
\label{p_s_p_f_relation}
v_{\kappa+1}&= p_s= p_f< v_{\kappa}\ \ \  ({\rm if}\ \kappa=1),  \nonumber\\ 
v_{\kappa+1}&\leq p_s\leq p_f= v_{\kappa}\ \ \ ({\rm if}\ \kappa\geq 2). 
\end{align}
This implies that all the good is sold until the next iteration from the iteration where bidders begin to clinch.
Suppose that $p_s<p_f$. For $p\in [p_s,p_f)$, it holds $|A(p)|=\kappa$ 
and thus 
\[
S(p)=\sum_{j\in A(p)\setminus i}b_j(p)/p=(\kappa-1)b_i(p)/p
\] for each $i\in [\kappa]$.
Therefore, the amount of remaining good is characterized by  
\begin{align}
\label{SP}
S(p)&=
\begin{cases}
1& {\rm if}\ p<p_s,\\
(\kappa-1) b_i(p)/p &{\rm if}\ p_s\leq p< p_f, \\
0& {\rm if}\ p\geq p_f.
\end{cases}
\end{align}

Using the above arguments, we present the main result of this section.
As a first step, we introduce the concept of {\it wishful allocation} \citep{GMP2020}, which represents the maximum possible amount of the good that each bidder can acquire at the end, given the current allocation and payment.
\begin{definition}[\citet{GMP2020}]
\label{derivative}
The wishful allocation $\Psi_i(p):\mathbf R_+\to \mathbf R_+$  of bidder $i$ at price $p$ is defined by 
$\Psi_i(p):=x_i(p)+b_i(p)/p.$
It is continuous and right-differentiable for all $p$, and the right-derivative is given by~$\partial_p \Psi_i(p)=-b_i(p)/p^2$.
\end{definition}
Note that $\Psi_i(p)$ is uniquely determined for each $p$. 
For example, in line 7, if bidder $i$ clinches $\delta_i$ unit, then $x_i(p)$ increases~by $\delta_i$ and $b_i(p)/p$ decreases by $\delta_i$, meaning that $\Psi_i(p)$ is unchanged.

We now present the following theorem, which provides an explicit formula for the functions in Algorithm~\ref{alg:algorithm}. 
Combining this with Theorems \ref{properties_sec5} (iii),  the execution of Algorithm \ref{alg:algorithm} is completely characterized by the triple $(p_s, p_f, \kappa)$. 
This result will be used in Section 4 to compare outcomes before and after the addition of a new bidder.

\begin{theorem}
\label{psi}
In Algorithm~\ref{alg:algorithm} under the symmetric setting, 
if $p_s<p_f$, it holds for each $i\in [\kappa]$~that 
\begin{align*}
x_i(p)&=\frac{1}{\kappa}-\frac{(\kappa-1)(\kappa \beta-p_s) (p_s)^{\kappa-1}p^{-\kappa}}{\kappa}\quad (p_s\leq p< p_f),\\
b_i(p)&=(\kappa \beta-p_s) (p_s)^{\kappa -1} p^{-(\kappa -1)}\ \ \qquad\qquad(p_s\leq p<p_f), \\
\Psi_i(p)&=\frac{1}{\kappa}+\frac{(\kappa \beta-p_s) (p_s)^{\kappa-1}p^{-\kappa}}{\kappa}\qquad\qquad (p_s\leq p\leq p_f).
\end{align*} 
\end{theorem}
The following example helps to understand the intuition of this formula: 
\begin{example}
\label{3}
Consider a multi-unit auction with $n=2$. Bidder 1 has a valuation of $4$ and bidder 2 has a valuation of $3$. 
Also, both bidders have equal budgets of $1$. At $p=1$, both bidders begin to clinch the good, 
and at $p=3$, bidder 2 drops out and all the remaining good is allocated to bidder 1. Then, by (\ref{p_s_p_f_relation})
and Definition \ref{interval}, we have $(p_s, p_f,\kappa)=(1, 3, 2)$. By Theorems \ref{properties_sec5} (iii) and \ref{psi}, we have $x^{\rm f}_1=\Psi_1(3)=1/2+1/(2\cdot 3^2)=5/9$ and $x^{\rm f}_2=x_2(3)=1/2-1/(2\cdot 3^2)=4/9$. Similarly, we have $\pi^{\rm f}_1=\beta=1$ and $\pi^{\rm f}_2=\beta-b_2(3)=1-1/3=2/3$.
\end{example}

\begin{proof}
First, we show $b_i(p_s)=\kappa \beta-p_s$ for each $i\in [\kappa]$. Combining $p_s<p_f$ with (\ref{p_s_p_f_relation}) and Lemma~\ref{p_s_p_f_common}, we have $(\kappa-1)\beta\leq p_s<p_f=v_{\kappa}$.
Then, we have $(\kappa-1) \beta\leq v_{\kappa+1}<v_{\kappa}$ or $v_{\kappa+1}<(\kappa-1) \beta<v_{\kappa}$.
In the first case, by Lemma~\ref{p_s_p_f_common}, it holds $p_s=v_{\kappa+1}$.
At $p=v_{\kappa+1}(=p_s)$, the number of active bidders is reduced to $\kappa$.
Then, in line 6, we have $\delta_i=S(v_{\kappa+1}-)-\sum_{j\in A(v_{\kappa+1})}b_j(v_{\kappa+1}-)/v_{\kappa+1}=1-(\kappa-1) \beta/p_s\geq 0$ for each $i\in [\kappa]$.
Since the transaction price is $p_s$ per unit, we have $b_i(p_s)=\beta-p_s\bigl(1-(\kappa-1)\beta/p_s\bigr)=\kappa \beta-p_s$.
In the second case, by Lemma~\ref{p_s_p_f_common}, it holds $p_s=(\kappa-1) \beta$.
Then, it holds $0\leq v_{\kappa+1}<(\kappa-1)\beta=p_s$.
At $p=p_s$, it holds $S(p_s)=(\kappa-1)\beta/p_s=~1$.
This means that no bidder clinches the good at price $p_s$, and thus we also have 
$b_i(p_s)=\beta=\kappa \beta-(\kappa-1) \beta=\kappa \beta-p_s$.

At $p\in [p_s,p_f)$, the remaining budget $b_i(p)$ of bidder $i\in [\kappa]$ is changed according to 
$\partial_p b_i(p)=- S(p)=-(\kappa-1)b_i(p)/p$ by (\ref{SP}).
Since $b_i(p)$ is continuous on $[p_s,p_f)$ and $b_i(p_s)=\kappa \beta-p_s$, 
solving the differential equation, we have 
\[
b_i(p)=(\kappa \beta-p_s) (p_s)^{\kappa -1} p^{-(\kappa -1)}\ (p_s\leq p<p_f).
\]

By Definitions \ref{interval} and \ref{derivative}, we have 
\[
\partial_p \Psi_i(p)=-b_i(p)/p^2=-(\kappa \beta-p_s) (p_s)^{\kappa -1} p^{-(\kappa+1)}\ \ {\rm and}\ \ 
\Psi_i(p_s)=x_i(p_s)+b_i(p_s)/p_s=\beta/p_s.
\]
Since $\Psi_i(p)$ is continuous for all $p$, we have 
\begin{align*}
\Psi_i(p)=\Psi_i(p_s)+\int^{p}_{p_s}-(\kappa \beta-p_s) (p_s)^{\kappa -1} p^{-(\kappa+1)}{\rm d}p
=\frac{1}{\kappa}+\frac{(\kappa \beta-p_s) (p_s)^{\kappa-1}p^{-\kappa}}{\kappa} \quad (p_s\leq p\leq p_f).
\end{align*}
Finally, by $x_i(p)=\Psi(p)-b_i(p)/p$ from Definition \ref{derivative}, we have 
\begin{align*}
x_i(p)=\frac{1}{\kappa}-\frac{(\kappa-1)(\kappa \beta-p_s) (p_s)^{\kappa-1}p^{-\kappa}}{\kappa}\ \ (p_s\leq p<p_f).
\end{align*}
\end{proof}
As an immediate corollary, we have the following:
\begin{corollary}
\label{common_x}
In Algorithm~\ref{alg:algorithm} under the symmetric setting, 
if $p_s<p_f$, it holds $x^{\rm f}_i>1/\kappa$ for each~$i\in[\kappa-1]$.
\end{corollary}
\begin{proof}
By Lemma \ref{p_s_p_f_common} and Theorem~\ref{psi}, we have $\Psi_i(p_f)>1/\kappa$ for each~$i\in [\kappa-1]$.
By $p_s<p_f$ and (\ref{p_s_p_f_relation}), we have $\kappa\geq 2$. Therefore, by Theorem~\ref{properties_sec5}~(iii), we have $x^{\rm f}_i(p)=\Psi_i(p_f)>1/\kappa$.\end{proof}

\section{Monotonicity as the Number of Bidders Increases}
This section covers a scenario where a new bidder $\theta$ is added to the input. We assume $v_{\theta} \neq v_i$ for each $i \in N$. The new input is indicated as $(N^\theta, v^\theta, B^\theta)$, where $N^\theta := N \cup \{\theta\}$, $v^\theta := v \cup \{v_\theta\}$, and $B^\theta := B \cup \{B_\theta\}$. In executing Algorithm~\ref{alg:algorithm}, bidder $\theta$ is placed so that bidders are in descending order of their valuations. Define $(x^{\rm f}, \pi^{\rm f}) := {\rm Adp}(N, v, B)$ and $(\tilde{x}^{\rm f}, \tilde{\pi}^{\rm f}) := {\rm Adp}(N^\theta, v^\theta, B^\theta)$. In the following, we compare these final outcomes to establish their relationship.

Algorithm~\ref{alg:algorithm} can be used in asymmetric settings. However, the addition of a new bidder with a different valuation and budget can significantly alter the outcome, making the comparison of outcomes highly complex.\footnote{The difficulty of these situations is discussed in Section 6.} Therefore, we consider the symmetric setting, where all bidders have equal budgets $\beta$, i.e., $B_i = \beta$ for each $i \in N^\theta$. Our main result in this section is as follows:

\begin{theorem}
\label{monotonicity}
In Algorithm~\ref{alg:algorithm} under the symmetric setting, the following relationship holds:
\begin{itemize}
\item[(i)] For each bidder $i\in N\setminus\kappa$, it holds $x^{\rm f}_i\geq \tilde{x}^{\rm f}_i$ and $\pi^{\rm f}_i\geq \tilde{\pi}^{\rm f}_i$.
\item[(ii)] It holds ${\rm LW}(x^{\rm f})\leq {\rm LW}(\tilde{x}^{\rm f})$ and ${\rm REV}(\pi^{\rm f}) \leq {\rm REV}(\tilde{\pi}^{\rm f})$.
\end{itemize}
\end{theorem}

Property (i) indicates that the monotonicity of allocation and payment holds for bidders in $N \setminus~\kappa$. 
Further, property~(ii) shows the monotonicity of LW and revenue as the number of bidders increases. Notably, this is the first study to address such monotonicity.
Since the adaptive clinching auction is the standard mechanism in budget-constrained auctions, Theorem~\ref{monotonicity} supports the use of LW as an efficiency measure in these auctions.

The following example complements Theorem~\ref{monotonicity}. It shows that SW is not necessarily monotone and that the outcome for bidder $\kappa$ may increase. 
\begin{example}
Consider the same market as in Example \ref{3}. By $x^{\rm f}_1 = 5/9$ and $x^{\rm f}_2 = 4/9$, the SW is $4 \cdot 5/9 + 3 \cdot 4/9 = 32/9 \approx 3.56$. Now consider a new bidder $\theta$ with a valuation of $2$ and a budget of $1$. When $p<~2$, no bidder can clinch the good. Then, at $p = 2$, bidder $\theta$ drops out of the auction, and bidders 1 and 2 each clinch 0.5 units of the good, paying 1. The SW is $3 \cdot 0.5 + 4 \cdot 0.5 = 3.5$. Consequently, the outcome for bidder $\kappa$ (i.e., bidder 2) might increase, and the SW is not monotone.
\end{example}

The rest of this section is dedicated to proving Theorem~\ref{monotonicity}. All notions (functions, sets, numbers, etc.) in ${\rm Adp}(N^\theta, v^\theta, B^\theta)$ are denoted by adding a tilde to the top of the original terms. For example, let $[\tilde{p}_s, \tilde{p}_f]$ represent the new clinching interval, and let $\tilde{\kappa}$ represent the new critical bidder. In Section 4.1, we identify the changes in the triple $(\tilde{p}_s, \tilde{p}_f, \tilde{\kappa})$ from $(p_s, p_f, \kappa)$. In Section~4.2, we analyze how the outcome changes, using Theorems~\ref{properties_sec5} (iii) and \ref{psi}.

\subsection{Changes of Clinching Interval}

According to (\ref{p_s_p_f_relation}), if $\tilde{p}_s$ can be determined, all the good is allocated until the next iteration. This helps us to determine $\tilde{p}_f$ and $\tilde{\kappa}$. Now we perform a case-by-case analysis, examining when bidders start clinching the good, and then calculating the triple of $(\tilde{p}_s, \tilde{p}_f, \tilde{\kappa})$. 
In all cases, we show that $(\tilde{p}_s, \tilde{p}_f, \tilde{\kappa})$ can be expressed by $v_\theta$ and $(p_s, p_f, \kappa)$ (with $p_f = v_\kappa$).

\begin{lemma}
\label{change_interval}
In ${\rm Adp}(N^\theta , v^\theta, B^\theta)$, it holds 
\begin{align*}
(\tilde{p}_s, \tilde{p}_f, \tilde{\kappa})= 
\begin{cases}
(p_s, p_f, \kappa) & \text{if } v_{\theta} \leq p_s, \\
(\min(v_{\theta}, v_1), \min(v_{\theta}, v_1), 1) & \text{if } \kappa = 1 
\text{ and } 
p_s < \min(v_{\theta}, v_1) < \beta, \\
(v_{\kappa}, \min(v_{\theta}, v_{\kappa-1}), \kappa) & \text{if } \kappa \geq 2\text{ and } 
p_s \leq v_{\kappa} < \min(v_{\theta}, \kappa \beta), \\
(v_{\theta}, v_{\kappa}, \kappa) & \text{if } \kappa \geq 2 \text{ and } 
 p_s < v_{\theta} < \min(v_{\kappa}, \kappa \beta), \\
(\kappa \beta, \min(v_{\theta}, v_{\kappa}), \kappa+1) & \text{if } p_s < \kappa \beta < \min(v_{\theta}, v_{\kappa}), \\
(\kappa \beta, \kappa \beta, \kappa+1) & \text{if } p_s < \kappa \beta = \min(v_{\theta}, v_{\kappa}).
\end{cases}
\end{align*}
\end{lemma}

In each case, the mechanism is executed as follows: 
\begin{itemize}[left=9pt]
\item The first case: Bidder $\theta$ drops out before any bidder begins to clinch~the good. 
The auction proceeds as if no bidder is~added. 
\item The second case: The entire good is allocated to the highest-valued bidder when the second-highest-valued bidder drops~out. 
\item The third case: No bidder clinches the good until bidder $\kappa$ drops out. Then, the bidders in $[\kappa-1]\cup \theta$ remain active and clinch~the good as the price increases. The next price where a bidder drops out is $\min(v_{\theta}, v_{\kappa-1})$. After that, all active bidders at that price clinch the good and exhaust their budgets. 
\item The fourth case: No bidder clinches the good until bidder $\theta$ drops out. Bidders in $[\kappa]$ remain active and clinch the~good as the price increases. When bidder $\kappa$ drops out, bidders in $[\kappa-1]$ clinch the good and exhaust their~budgets. 

\item The fifth case: No bidder clinches the good until the price reaches $\kappa \beta$. At this point, bidders in $[\kappa]\cup \theta$ are still active and start to clinch the good. The next price where a bidder drops out is $\min(v_{\theta}, v_{\kappa})$. Afterward, all active bidders clinch the good and exhaust their budgets. 
\item The sixth case: No bidder clinches the good until the price reaches $\kappa \beta$, at which either bidder $\theta$ or $\kappa$ drops out. Since each of the remaining active bidder has a demand of $1/\kappa$, all the good is allocated to them, exhausting their~budgets.
\end{itemize}
In Section 4.2, we refer to these cases as Case 1 through Case 6.

\subsection{Proof of Theorem \ref{monotonicity}}
For property (i), the outcomes $(x^{\rm f}, \pi^{\rm f})$ and $(\tilde{x}^{\rm f}, \tilde{\pi}^{\rm f})$ are compared. In some cases, detecting changes in outcomes requires comparing $\tilde{b}_i(p)$ with $b_i(p)$ and $\tilde{\Psi}_i(p)$ with $\Psi_i(p)$. Using $(\tilde{p}_s, \tilde{p}_f, \tilde{\kappa})$ from Lemma~\ref{change_interval}, 
Theorem~\ref{psi} can be applied to derive explicit formula for 
$\tilde{b}_i(p)$ and $\tilde{\Psi}_i(p)$ for each $i \in [\kappa]$ 
since all bidders including bidder $\theta$ have equal budgets. 
For property (ii), we use the following lemma. It shows that LW and revenue can be expressed simply in the symmetric setting:

\begin{lemma}
\label{common_LW_pi}
Let $(x^{\rm f}, \pi^{\rm f})$ be the outcome of Algorithm~\ref{alg:algorithm} 
in the symmetric setting and $\kappa$ be the critical bidder in Definition \ref{critical}. Then, it holds 
\begin{align*}
{\rm LW}(x) &= (\kappa - 1)\beta + \min(v_\kappa x^{\rm f}_\kappa, \beta) \leq \kappa \beta,\\
{\rm REV}(\pi) &= (\kappa - 1)\beta + \pi^{\rm f}_\kappa \leq \kappa \beta.
\end{align*} 
\end{lemma}
\begin{proof}
If $\kappa=1$, by Theorem \ref{properties_sec5} (iii), the claim trivially holds. Suppose that $\kappa\geq 2$. 
By budget-feasibility and IR (Theorem~\ref{properties_sec5}~(i)), it holds $v_i x^{\rm f}_i\geq \pi^{\rm f}_i$ and $\beta\geq \pi^{\rm f}_i$ for each $i\in N$.
By Theorem~\ref{properties_sec5}~(iii), it holds $\pi^{\rm f}_i=\beta$ for each $i\in[\kappa-1]$ 
and $(x^{\rm f}_i,\pi^{\rm f}_i)=(0,0)$ for each $i>\kappa$.
This means that, for each $i\in[\kappa-1]$, it holds $\min(v_i x^{\rm f}_i,\beta)=\pi^{\rm f}_i=\beta$ 
and for each $i>\kappa$, it holds $\min(v_i x^{\rm f}_i,\beta)=\pi^{\rm f}_i=0$.
Therefore, we have 
\begin{align*}
{\rm LW}(x^{\rm f})&=(\kappa-1) \beta+\min(v_{\kappa} x^{\rm f}_{\kappa}, \beta)\leq\kappa\beta\\
{\rm REV}(\pi^{\rm f})&=(\kappa-1)\beta+\pi^{\rm f}_{\kappa}\leq \kappa\beta,
\end{align*}
where the last inequality holds by budget-feasibility.
\end{proof}

We perform a case-by-case analysis based on the new clinching interval. 
The fourth and fifth cases require a precise comparison of functions using Theorem \ref{psi}. 
\begin{proof}[Proof of Theorem \ref{monotonicity}]
In all cases of Lemma \ref{change_interval}, it holds $\tilde{\kappa}\geq \kappa$.
By Theorem~\ref{properties_sec5}~(iii), we have 
\begin{equation}
\label{kappa+1}
x^{\rm f}_i=\tilde{x}^{\rm f}_i=\pi^{\rm f}_i=\tilde{\pi}^{\rm f}_i=0\ \ (i=\kappa+1,\kappa+2,\ldots, n),
\end{equation}
from which property (i) holds trivially for these bidders.

Case 1: $v_{\theta}<p_s$. Since bidder $\theta$ drops out before the price reaches $p_s$, 
the outcome is unchanged.
Then, all the inequalities in Theorem \ref{monotonicity} hold in equality.

Case 2: $\kappa=1$ and $p_s< \min(v_{\theta}, v_1)<\beta$.
By $(\ref{kappa+1})$, property (i) trivially holds.
In this case, by Lemma \ref{change_interval}, it holds 
$(\tilde{p}_s,\tilde{p}_f, \tilde{\kappa})=(\min(v_{\theta}, v_1), \min(v_{\theta}, v_1), 1)$.
By $\tilde{\kappa}=\kappa=1$ and Theorem \ref{properties_sec5} (iii), under both inputs, 
the highest-valued bidder clinches all the good and pays the second highest value.
Thus, property (ii) holds by 
\begin{align*}
{\rm LW}(x^{\rm f})&=\min(v_1, \beta)\leq \min(\max(v_1,v_\theta), \beta)={\rm LW}(\tilde{x}^{\rm f}),\\
{\rm REV}(\pi^{\rm f})&=v_2=p_s<\min(v_{\theta}, v_1)={\rm REV}(\tilde{\pi}^{\rm f}).
\end{align*}

Case 3:  $\kappa\geq 2$ and $p_s\leq v_{\kappa}<\min(v_{\theta},\kappa \beta)$. 
By Lemma \ref{change_interval}, it holds $(\tilde{p}_s,\tilde{p}_f,\tilde{\kappa})=(v_{\kappa}, \min(v_{\theta},v_{\kappa-1}), \kappa)$.
Suppose that  bidder $\theta$ drops out before the dropping of bidder $\kappa-1$.
In the execution of Algorithm~\ref{alg:algorithm}, each good is sold at the price less than $v_{\kappa}(=p_f)$ under $(N, v, B)$ and more than $v_{\kappa}(=\tilde{p}_s)$ under $(N_\theta, v_\theta, B_\theta)$.
Then, we have $\tilde{x}^{\rm f}_i\leq \tilde{\pi}^{\rm f}_i/\tilde{p}_s=\beta/p_f\leq x^{\rm f}_i$ for each $i\in [\kappa-1]$.
By budget-feasibility and Theorem~\ref{properties_sec5}~(iii), it holds $\tilde{\pi}^{\rm f}_i\leq \beta=\pi^{\rm f}_i$. 
Combining them with (\ref{kappa+1}), we have property (i).

Since all the good is sold in the mechanism from Theorem~\ref{properties_sec5}~(ii), 
by property (i), we have
$x^{\rm f}_{\kappa}=1-\sum_{i\in [\kappa-1]}x^{\rm f}_{i}\leq 1-\sum_{i\in [\kappa-1]}\tilde{x}^{\rm f}_{i}=\tilde{x}^{\rm f}_{\theta}$.
Note that $\tilde{x}^{\rm f}_\kappa=0$ holds by $\tilde{p}_s=v_\kappa$. 
Combining this with $p_f\leq\tilde{p}_s$, we have 
$\pi^{\rm f}_{\kappa}\leq p_f x^{\rm f}_{\kappa}=\tilde{p}_s x^{\rm f}_{\kappa}\leq \tilde{p}_s \tilde{x}^{\rm f}_{\theta}\leq \tilde{\pi}^{\rm f}_{\theta}$.
Thus, property~(ii) holds by 
\begin{align*}
{\rm LW}(x^{\rm f})&=(\kappa-1) \beta+\min(v_{\kappa} x^{\rm f}_{\kappa}, \beta)
\leq (\kappa-1) \beta+\min(v_{\theta} \tilde{x}^{\rm f}_{\theta}, \beta)={\rm LW}(\tilde{x}^{\rm f}),\\
{\rm REV}(\pi^{\rm f})&=(\kappa-1)\beta+\pi^{\rm f}_{\kappa}\leq (\kappa-1)\beta+\tilde{\pi}^{\rm f}_{\theta}={\rm REV}(\tilde{\pi}^{\rm f}).
\end{align*}
Suppose that bidder $\kappa-1$ drops out before bidder $\theta$. 
By changing the role of $\kappa-1$ and $\theta$, we also have properties (i) and (ii).

Case 4: $\kappa\geq 2$ and $p_s< v_{\theta}<\min(v_{\kappa}, \kappa B)$.  
By Lemma \ref{change_interval}, it holds $(\tilde{p}_s,\tilde{p}_f, \tilde{\kappa})=(v_{\theta}, v_{\kappa}, \kappa)$.
By $\tilde{p}_s<\tilde{p}_f$, substituting them into Theorem \ref{psi}, we have 
\begin{align*}
\tilde{b}_i(p)&=(\kappa \beta-v_{\theta}) v_{\theta}^{\kappa -1} p^{-(\kappa -1)}\ \ \ \ \, (\tilde{p}_s\leq p< \tilde{p}_f), \\
\tilde{\Psi}_i(p)&=\frac{1}{\kappa}+\frac{(\kappa \beta-v_{\theta}) v_{\theta}^{\kappa-1}p^{-\kappa}}{\kappa}\ \ \, (\tilde{p}_s\leq p\leq \tilde{p}_f), 
\end{align*}
for each $i\in [\kappa]$.
Observe that $\tilde{b}_i(p)$ and $\tilde{\Psi}_i(p)$ are obtained from $b_i(p)$ and $\Psi_i(p)$, respectively, by replacing $p_s$ 
with $v_{\theta}$.
Note that by $p_s<v_\kappa=p_f$, we can also apply Theorem~\ref{psi} for $(p_s, p_f, \kappa)$. To compare these functions, define a function 
$G_{\kappa}(x):=(\kappa \beta-x)x^{\kappa-1}\ ((\kappa-1) \beta<x<\kappa \beta)$.
Then, by $\kappa\geq 2$, we have $G_{\kappa}'(x)=\kappa x^{\kappa-2}\bigl((\kappa-1)\beta-x\bigr)\leq~0$
for $(\kappa-1) \beta<x<\kappa \beta$ and thus $G_{\kappa}(x)$ is weakly decreasing on~$x$.
By $p_s<v_{\theta}$, it holds $G_{\kappa}(v_\theta)\leq G_{\kappa}(p_s)$ and thus 
\begin{align*}
\tilde{b}_i(p)&\leq b_i(p)\ \ \, {\rm for}\ \ \tilde{p}_s\leq p<\tilde{p}_f,\\
\tilde{\Psi}_i(p)&\leq \Psi_i(p)\ \ {\rm for}\ \ \tilde{p}_s\leq p\leq \tilde{p}_f.
\end{align*}
 By $\tilde{\kappa}=\kappa\geq 2$ and Theorem \ref{properties_sec5} (iii),  it holds for each $i\in [\kappa-1]$ that 
$x^{\rm f}_i=x_i(p_f-)+b_i(p_f-)/p_f=\Psi_i(p_f)$ and $\tilde{x}^{\rm f}_i=\tilde{x}_i(\tilde{p}_f-)+\tilde{b}_i(\tilde{p}_f-)/\tilde{p}_f=\tilde{\Psi}_i(p_f)$.
Then, by $p_f=\tilde{p}_f=v_{\kappa}$, we have 
$\tilde{x}^{\rm f}_i=\tilde{\Psi}_i(\tilde{p}_f)\leq \Psi_i(p_f)=x^{\rm f}_i$.
By Theorem~\ref{properties_sec5}~(iii), we have $\pi^{\rm f}_i=\beta=\tilde{\pi}^{\rm f}_i$ 
for each $i\in [\kappa-1]$.
Combining this with (\ref{kappa+1}), property (i)~holds.

For property (ii), we consider the allocation of bidder $\kappa$. 
Remark that by $\tilde{p}_s=v_\theta$, it holds 
$\tilde{x}^{\rm f}_\theta=\tilde{\pi}^{\rm f}_\theta=0$.
Since all the good is sold by Theorem~\ref{properties_sec5}~(ii), 
we have 
\[
x^{\rm f}_{\kappa}=1-\sum_{i\in N\setminus \kappa} x^{\rm f}_{i}
=1-\sum_{i\in N\setminus \kappa} x^{\rm f}_{i}-\tilde{x}^{\rm f}_\theta
\leq 1-\sum_{i\in N^\theta\setminus \kappa}\tilde{x}^{\rm f}_{i}=\tilde{x}^{\rm f}_{\kappa},
\] 
where the inequality holds by property (i).
By Theorem~\ref{properties_sec5}~(iii) and $p_f=\tilde{p}_f$, we have  
$\tilde{\pi}^{\rm f}_{\kappa}=\beta-\tilde{b}_{\kappa}(\tilde{p}_f-)\geq \beta-b_{\kappa}(p_f-)=\pi^{\rm f}_{\kappa}$.
Therefore, by Lemma~\ref{common_LW_pi} and $\kappa=\tilde{\kappa}$, property (ii) holds by 
\begin{align*}
{\rm LW}(x^{\rm f})&=(\kappa-1)\beta+\min(v_{\kappa} x^{\rm f}_{\kappa}, \beta)\leq(\tilde{\kappa}-1)\beta+\min(v_{\kappa} \tilde{x}^{\rm f}_{\kappa}, \beta)={\rm LW}(\tilde{x}^{\rm f}),\\
{\rm REV}(\pi^{\rm f})&=(\kappa-1)\beta+\pi^{\rm f}_{\kappa}\leq(\tilde{\kappa}-1)\beta+\tilde{\pi}^{\rm f}_{\kappa}={\rm REV}(\tilde{\pi}^{\rm f}).
\end{align*}

Case 5: $p_s<\kappa \beta<\min(v_{\theta}, v_{\kappa})$. 
If $\kappa=1$, then property (i) holds by (\ref{kappa+1}). Suppose that $\kappa\geq 2$.
By Lemma \ref{change_interval}, we have $(\tilde{p}_s,\tilde{p}_f,\tilde{\kappa})=(\kappa \beta, \min(v_{\theta}, v_{\kappa}),\kappa+1)$.
By $\tilde{p}_s<\tilde{p}_f$, substituting them into Theorem \ref{psi}, we have 
\[
\tilde{\Psi}_i(p)=\frac{1}{\kappa+1}+\frac{\beta (\kappa \beta)^{\kappa}p^{-(\kappa+1)}}{\kappa+1}\quad (\tilde{p}_s\leq p\leq \tilde{p}_f).
\]
At $p=\tilde{p}_f>\tilde{p}_s$,
bidder $\theta$ or $\kappa$ is dropping.
By $\tilde{\kappa}=\kappa+1\geq 2$ and Theorem~\ref{properties_sec5}~(iii), 
it holds $\tilde{x}^{\rm f}_i=\tilde{\Psi}_i(\tilde{p}_f)$ 
for each $i\in [\kappa-1]$.
Then, by $\tilde{p}_f>\tilde{p}_s=\kappa\beta$, 
we also have 
$\tilde{x}^{\rm f}_i=\tilde{\Psi}_i(\tilde{p}_f) \leq \tilde{\Psi}_i(\kappa \beta) =1/(\kappa+1)+1/\kappa(\kappa+1)=1/\kappa< x^{\rm f}_i$,
where the last inequality holds by Corollary \ref{common_x}.
By Theorem~\ref{properties_sec5}~(iii), it holds $\tilde{\pi}^{\rm f}_i=\beta=\pi^{\rm f}_i$ for each $i\in [\kappa-1]$.
Combining them with (\ref{kappa+1}), we have property~(i).

For property (ii), by Lemma \ref{common_LW_pi} and $\tilde{\kappa}=\kappa+1$, we have  
\[
{\rm LW}(x^{\rm f})\leq \kappa \beta=(\tilde{\kappa}-1) \beta\leq {\rm LW}(\tilde{x}^{\rm f})\ \  {\rm and}\  \ 
{\rm REV}(\pi^{\rm f})\leq \kappa \beta=(\tilde{\kappa}-1) \beta\leq {\rm REV}(\tilde{\pi}^{\rm f}).
\]

Case 6:  $p_s<\kappa \beta=\min(v_{\theta}, v_{\kappa})$.  
If $\kappa=1$, then property (i) holds by (\ref{kappa+1}). Suppose that $\kappa\geq 2$.
By Lemma \ref{change_interval}, it holds 
$(\tilde{p}_s,\tilde{p}_f, \tilde{\kappa})=(\kappa \beta, \kappa \beta, \kappa+1)$.
Suppose that bidder $\theta$ drops out before the dropping of bidder $\kappa$.
Consequently, by $\tilde{p}_s=\tilde{p}_f$, all the good is allocated at $p=\kappa \beta$. 
By $\tilde{\kappa}=\kappa+1\geq 2$ and Theorem~\ref{properties_sec5}~(iii), 
it holds $\tilde{\pi}^{\rm f}_i=\beta$ and $\tilde{x}^{\rm f}_i=\beta/(\kappa \beta)=1/\kappa$ for each $i\in [\kappa]$.
By $p_s<v_\kappa=p_f$ and Corollary \ref{common_x}, we have  
$x^{\rm f}_i> 1/\kappa=\tilde{x}^{\rm f}_i$ for each $i\in [\kappa-1]$.
By Theorem~\ref{properties_sec5}~(iii), it also holds $\pi^{\rm f}_i=\beta=\tilde{\pi}^{\rm f}_i$.
Combining them with (\ref{kappa+1}), property (i) also holds.
Suppose that  bidder $\kappa$ drops out before bidder $\theta$.
Then, by changing the role of $\kappa$ and $\theta$, we also have property~(i). 

Property (ii) holds by the same arguments as in Case~5.
\end{proof}

As an immediate corollary, we can also provide the guarantee of the ratio between the LW values before and after the input change. The same is true if we replace LW with revenue.
\begin{corollary}
It holds $1\leq {\rm LW}(\tilde{x}^{\rm f})/{\rm LW}(x^{\rm f})\leq (\kappa+1)/(\kappa-1)$.
\end{corollary}
\begin{proof}
By Theorem \ref{monotonicity}, it holds 
${\rm LW}(\tilde{x}^{\rm f})/{\rm LW}(x^{\rm f})\geq 1$. 
By $\tilde{\kappa}\leq \kappa+1$ and Lemma \ref{common_LW_pi}, we have ${\rm LW}(\tilde{x}^{\rm f})/{\rm LW}(x^{\rm f})\leq (\kappa+1)/(\kappa-1)$. 
\end{proof}

\section{Online Arrival}
The remainder of this paper investigates the scope of the monotonicity result in Theorem~\ref{monotonicity}.
We now consider another symmetric setting in which new bidders, denoted by 
$\Theta := \{\theta_1, \theta_2, \ldots, \theta_t\}$, arrive online in Algorithm~\ref{alg:algorithm}.  
Define $N^\Theta := N \cup \Theta$, $v^\Theta := v \cup \{v_\theta\}_{\theta\in\Theta}$, and $B^\Theta := B \cup \{B_\theta\}_{\theta\in\Theta}$. All other notations follow the ones introduced in Section 4.

To formalize this setting, we apply the following conditions to the arrival process:
\begin{itemize}[left=5pt]
\item Every new bidder $\theta_k$ arrives at $p=\gamma_k$ and immediately reports their bid $v'_{\theta_k}\, (>\gamma_k)$. They are added in line~11 of Algorithm~\ref{alg:algorithm} if $\gamma_k=v'_j$ for some $j\in N$, and in line~3 otherwise.
\item New bidders arrive sequentially in ascending order of arrival prices, i.e., 
$\gamma_1<\gamma_2<\ldots<\gamma_t$.
\item All bidders in $N^{\Theta}$ have equal budgets, i.e., $B_i=\beta\ (i\in N^{\Theta})$.
\item The auctioneer has no prior knowledge of $\Gamma:=\{\gamma_1,\gamma_2,\ldots,\gamma_t\}$.
\end{itemize}

It is straightforward to verify that Algorithm~\ref{alg:algorithm} can accommodate the online arrival of bidders after modifying the definition of $A(p)$ as follows:
\[
A(p) = \{i \in N \mid p < v_i\} \cup \{\theta_k \in \Theta \mid \gamma_k < p < v_{\theta_k} \}.
\]
In particular, the algorithm still satisfies IC and IR. 
This is because bids are only used to 
determine when bidders drop out, and no bidder clinches the good at a price higher than their bid.  
Based on this, we assume that bidders report their valuations truthfully as their bids.

Our main result in this section is to show that Theorem~\ref{monotonicity} extends to this online setting.
\begin{theorem}
\label{onlinesupply}
Theorem~\ref{monotonicity} remains valid even when new bidders in $\Theta$ arrive online.
\end{theorem}

Theorem~\ref{onlinesupply} lays the groundwork for extending clinching auctions to social networks, where bidders arrive online, as in the mechanism proposed by \citet{XSK2022}.  
It implies that, in symmetric settings, the addition of bidders through information diffusion improves efficiency and revenue, a desirable property for mechanisms in social networks.

In the remainder of this section, we show Theorem \ref{onlinesupply}.
The analysis can be limited to $p_s\leq \gamma_1$ and $p_s\neq p_f$ without loss of generality.\footnote{If there exist new bidders who arrive before the price reaches $p_s$, 
as no goods have been allocated at their arrival, 
these bidders can be treated as initial bidders by (repeatedly) applying Theorem~\ref{monotonicity}.
Moreover, if $p_s=p_f$, all the goods are already allocated at the moment of bidder $\theta_1$'s arrival.}
Then, we have $\kappa\geq 2$ by Theorem~\ref{properties_sec5}~(iii).

Just when bidder $\theta_1$ arrives at $p=\gamma_1$, it holds 
\begin{align*}
x_i(\gamma_1)=\tilde{x}_i(\gamma_1),\ \ b_i(\gamma_1)=\tilde{b}_i(\gamma_1)<\beta
=\tilde{b}_\theta(\gamma_1),\ {\rm and }\ \Psi_i(\gamma_1)=\tilde{\Psi}_i(\gamma_1)
\end{align*}
for each $i\in [\kappa]$.
Also, by $v_{\kappa+1}\leq p_s<\gamma_1$ from Lemma \ref{p_s_p_f_common}, we have
\begin{align}
\label{x_pi}
x^{\rm f}_i=\tilde{x}^{\rm f}_i=\pi^{\rm f}_i=\tilde{\pi}^{\rm f}_i=0\ \ (i=\kappa+1,\kappa+2,\ldots, n).
\end{align}
As certain goods have already been allocated to bidders in $[\kappa]$, 
the relationship between each bidder $i$'s demand and the number of remaining good at $p=\gamma_1$ is given by:

\begin{align}
\label{after_arrival}
S(\gamma_1)&=\tilde{S}(\gamma_1)=\sum_{j\in [\kappa] \setminus i}b_j(\gamma_1)/\gamma_1 
<\sum_{j\in [\kappa]}b_j(\gamma_1)/\gamma_1
<\sum_{j\in [\kappa]\cup\theta \setminus i} b_j(\gamma_1)/\gamma_1.
\end{align}

Immediately after the arrival of $\theta_1$, no bidder in $[\kappa] \cup \{\theta_1\}$ belongs to the clinching set.  
This implies that a bidder who was previously in the clinching set may exit due to the arrival of new bidders,  
which contrasts with the offline setting.

Furthermore, since new bidders have more remaining budgets than initial bidders at the moment of their arrival, 
active bidders are somewhat asymmetric. For instance, if an initial bidder enters the clinching set 
while there exists an active bidder $\theta\in \Theta$, then $\theta$ must also belong to the clinching set.
However, the converse does not necessarily hold. 

By such arguments, we establish that even with the online arrival of new bidders,  
the mechanism preserves many useful properties, as stated in Theorem~\ref{properties_sec5}.
\begin{lemma}
\label{dropping_sec5}
Consider Algorithm~\ref{alg:algorithm} under the symmetric setting with new bidders arriving online.  
New bidders $\Theta$ sequentially enter the auction, and each bidder in $N \cup \Theta$ has an equal budget $\beta$.  
Then, the following properties hold:
\begin{itemize}
\item[{\rm (i)}] For each $i\in N^\Theta$, it holds $\min(v_i \tilde{x}^{\rm f}_i,\beta)\geq \tilde{\pi}^{\rm f}_i$.
\item[{\rm (ii)}] At the end of the auction, it holds $\sum_{i\in N^\Theta}\tilde{x}^{\rm f}_i=1$, i.e., all the good is allocated.
\item[{\rm (iii)}] All the remaining good is allocated to active bidders at the first price $p$ where a bidder in the clinching set is dropping, i.e., $\tilde{p}_f=p$. Then, it holds $\tilde{x}^{\rm f}_i=\tilde{\Psi}_i(\tilde{p}_f)$ and $\tilde{\pi}^{\rm f}_i=\beta$ for each $i\in \tilde{A}(\tilde{p}_f)$, 
and $\tilde{x}^{\rm f}_i=\tilde{x}_i(\tilde{p}_f-)$ and $\tilde{\pi}^{\rm f}_i=\beta-\tilde{b}_i(\tilde{p}_f-)$ for 
bidder $i$ with $v_i=\tilde{p}_f$.
\end{itemize}
\end{lemma}

We are ready to prove Theorem \ref{onlinesupply}.
\begin{proof}[Proof of Theorem \ref{onlinesupply}]
By budget feasibility and Theorem~\ref{properties_sec5}~(iii), we have 
$\tilde{\pi}^{\rm f}_i\leq \beta=\pi^{\rm f}_i$ for each $i\in [\kappa-1]$. 
Combining this with \eqref{x_pi}, yields the monotonicity of payment.  

To establish the remaining claims, we conduct a case-by-case analysis  
based on which bidder clinches the good in the price interval $[\gamma_1, v_\kappa)$. 
The following cases exhaust all possible scenarios:

Case 1: No bidders in $[\kappa]$ clinches the good.
For property (i), the key observation is that 
bidders in $[\kappa-1]$ clinch the good at a price higher than 
$v_{\kappa}(=p_f)$ after the arrival of new bidders under $(N^\Theta, v^\Theta, B^\Theta)$, 
while at a price lower than $v_{\kappa}$ under $(N, v, B)$.
Then, it implies 
\[
\tilde{x}^{\rm f}_i\leq \tilde{x}_i(\gamma_1)+\tilde{b}_i(\gamma_1)/v_{\kappa}=x_i(\gamma_1)+b_i(\gamma_1)/v_{\kappa}\leq x^{\rm f}_i
\]
for each $i\in [\kappa-1]$. 

For property (ii), we again use the observation. The payments are bounded as follows: 
\begin{align}
\label{payment_online}
\sum_{i\in [\kappa-1]}\pi^{\rm f}_i&=
\sum_{i\in[\kappa-1]} (\beta-b_i(\gamma_1))+\sum_{i\in [\kappa-1]}\bigl(\pi^{\rm f}_i- (\beta-b_i(\gamma_1))\bigr)\nonumber\\
&\leq \sum_{i\in[\kappa-1]} (\beta-b_i(\gamma_1))+v_\kappa \bigl(1-\sum_{i\in [\kappa]}x_i(\gamma_1)\bigr)
\leq \sum_{i\in [\kappa-1]\cup\Theta}\tilde{\pi}^{\rm f}_i,
\end{align}
where $\beta-b_i(\gamma_1)$ represents the current payment of bidder $i$ at $p=\gamma_1$ 
and $\pi^{\rm f}_i- (\beta-b_i(\gamma_1))$ represents their remaining payment afterward. 
Then, by (\ref{x_pi}) and Lemma \ref{dropping_sec5} (i), we have 
\begin{align*}
{\rm LW}(x^{\rm f})&=\min(v_{\kappa} x^{\rm f}_{\kappa}, \beta)+\sum_{i\in [\kappa-1]}\pi^{\rm f}_i
\leq\min(v_{\kappa} \tilde{x}^{\rm f}_{\kappa}, \beta)+\sum_{i\in [\kappa-1]\cup\Theta}\tilde{\pi}^{\rm f}_i\leq {\rm LW}(\tilde{x}^{\rm f}),
\\
{\rm REV}(\pi^{\rm f})&= \pi^{\rm f}_\kappa+\sum_{i\in [\kappa-1]}\pi^{\rm f}_i
\leq \tilde{\pi}^{\rm f}_\kappa+\sum_{i\in [\kappa-1]}\tilde{\pi}^{\rm f}_i
={\rm REV}(\tilde{\pi}^{\rm f}).
\end{align*}

Case 2: Only bidders in $[\kappa]$ clinch the good.
This case occurs only when new bidders repeatedly arrive at $[\gamma_1, v_\kappa)$ 
and drop out without entering the clinching set.
This follows from the fact that when bidders in $[\kappa]$ clinch the good,  
any active new bidder must also clinch it due to their larger remaining budgets.  
We distinguish two cases depending on whether the bidders in $[\kappa]$ 
are in the clinching set at $p=v_\kappa$.

Case 2-1: Bidders in $[\kappa]$ are in the clinching set just~before~$p=v_\kappa$. 
For property (i), we extend Theorem \ref{psi} to obtain the following relations:
\begin{claim}
\label{tildepsi_relation}
For each $i\in [\kappa]$, it holds 
$\tilde{\Psi}_i(v_{\kappa})\leq \Psi_i(v_{\kappa})\ \  {\rm and}\ \  \tilde{b}_i(v_{\kappa}-)\leq~b_i(v_{\kappa}-)$.
\end{claim}
The proof is given in Appendix C.4.
Combining this with Theorem \ref{properties_sec5} and Lemma \ref{tildepsi_relation}, for each $i\in [\kappa-1]$, 
the allocation monotonicity holds by $\tilde{x}^{\rm f}_i=\tilde{\Psi}_i (v_\kappa)\leq \Psi_i (v_\kappa)=x^{\rm f}_i$.
Since it~holds 
\begin{align*}
\tilde{x}^{\rm f}_\kappa=1-\sum_{i\in [\kappa-1]} \tilde{x}^{\rm f}_i\geq 1-\sum_{i\in [\kappa-1]} x^{\rm f}_i=x^{\rm f}_\kappa
\ \ {\rm and}\ \ 
\tilde{\pi}^{\rm f}_\kappa=\beta-\tilde{b}_\kappa(v_{\kappa}-)\geq \beta-b_\kappa(v_{\kappa}-)=\pi^{\rm f}_\kappa, 
\end{align*}
property (ii) holds by applying the same arguments as in Case~4 in the proof of Theorem \ref{monotonicity}.

Case 2-2: No bidder in $[\kappa]$ is in the clinching set immediately before $p=v_\kappa$.
This case occurs when bidders in $[\kappa]$ leave the clinching set by the arrival of a new bidder and then 
the price reaches $v_\kappa$. After the arrival, all the remaining good is sold 
at the price higher than $v_\kappa$. Now we apply the same argument in Case 1 based on Case 2-1.
Let $p_1$ be the price at which a new bidder who is active at $p=v_{\kappa}$ 
arrives (if multiple such prices exist, take the lowest one). 

Consider the case where no new bidders arrive in the price interval $[p_1,v_{\kappa})$ 
and bidders in $[\kappa]$ are in the clinching set just before $p=v_\kappa$. 
Let $\hat{x}^{\rm f}$ and $\hat{\pi}^{\rm f}$ be the final allocation in such situation.
Since this case is included in Case 2-1, we have 
$\hat{x}^{\rm f}_i\leq x^{\rm f}_i\ (i\in [\kappa-1])$, ${\rm LW}(\hat{x}^{\rm f})\geq {\rm LW}(x^{\rm f})$, and 
${\rm REV}(\hat{\pi}^{\rm f})\geq {\rm REV}(\pi^{\rm f})$.
Therefore, it suffices to show that 
$\tilde{x}^{\rm f}_i\leq \hat{x}^{\rm f}_i\ (i\in [\kappa-1])$, ${\rm LW}(\tilde{x}^{\rm f})\geq {\rm LW}(\hat{x}^{\rm f})$, and 
${\rm REV}(\tilde{\pi}^{\rm f})\geq {\rm REV}(\hat{\pi}^{\rm f})$.

For property (i), we compare $\tilde{x}^{\rm f}$ and $\hat{x}^{\rm f}$.
The key observation is that 
bidders in $[\kappa-1]$ clinch the good at a price higher than 
$v_{\kappa}(=p_f)$ after the arrival at $p_1$ in the former, 
while at a price lower than $v_{\kappa}$ in the latter.
Let $\hat{x}(p)$ and $\hat{b}(p)$ denote the functions of the allocation of goods and the remaining budget 
of bidder $i$ at price $p$, respectively. 
Then, it holds $\hat{x}_i(p_1)=\tilde{x}_i(p_1)=\tilde{x}^{\rm f}_\kappa\ \ {\rm and}\ \ \hat{b}_i(p_1)=\tilde{b}_i(p_1)=\beta-\tilde{\pi}^{\rm f}_\kappa$ for each $i\in [\kappa]$.
Thus, we have   
$\tilde{x}^{\rm f}_i\leq \tilde{x}_i(p_1)+\tilde{b}_i(p_1)/v_{\kappa}
=\hat{x}_i(p_1)+\hat{b}_i(p_1)/v_{\kappa}\leq \hat{x}^{\rm f}_i$
for each $i\in [\kappa-1]$. 

For property (ii), we again use the observation: 
Similar to (\ref{payment_online}), the payment is bounded by
\begin{align*}
&\max\bigl(v_\kappa (\hat{x}^{\rm f}_\kappa-\hat{x}_\kappa(p_1)), \hat{\pi}^{\rm f}_\kappa- (\beta-\hat{b}_\kappa(p_1))\bigr)
+\sum_{i\in [\kappa-1]}\bigl(\hat{\pi}^{\rm f}_i- (\beta-\hat{b}_i(p_1))\bigr)\\
&\qquad\quad\leq v_\kappa \bigl(1-\sum_{i\in [\kappa]}\hat{x}_i(p_1)\bigr)
=v_\kappa \bigl(1-\sum_{i\in [\kappa]}\tilde{x}_i(p_1)\bigr)
\leq \sum_{i\in [\kappa-1]\cup\Theta}\tilde{\pi}^{\rm f}_i-\sum_{i\in[\kappa-1]} (\beta-\hat{b}_i(p_1)),
\end{align*}
Consequently, by Lemma \ref{dropping_sec5} (i) and (\ref{x_pi}), we have 
\begin{align*}
 {\rm LW}(\hat{x}^{\rm f})&=\min(v_{\kappa} \hat{x}^{\rm f}_{\kappa}, \beta)
+\sum_{i\in [\kappa-1]}\hat{\pi}^{\rm f}_i
\leq\min(v_{\kappa} \tilde{x}^{\rm f}_{\kappa}, \beta)+\sum_{i\in [\kappa-1]\cup\Theta}\tilde{\pi}^{\rm f}_i
\leq {\rm LW}(\tilde{x}^{\rm f}),\\
{\rm REV}(\hat{\pi}^{\rm f})&= 
\hat{\pi}^{\rm f}_\kappa+\sum_{i\in [\kappa-1]}\hat{\pi}^{\rm f}_i
\leq \tilde{\pi}^{\rm f}_\kappa+\sum_{i\in [\kappa-1]\cup\Theta}\tilde{\pi}^{\rm f}_i
={\rm REV}(\tilde{\pi}^{\rm f}).
\end{align*}

Case 3: Other Cases.
There exists a new bidder who clinches the good. 
For property (i), we perform the case-by-case analysis on the relationship between $\tilde{p}_f$ and $v_\kappa$. 
In all cases, we show $\tilde{x}^{\rm f}_i\leq 1/\kappa$ for each $i\in [\kappa-1]$.
Combining this with Corollary \ref{common_x}, we have $\tilde{x}^{\rm f}_i<x^{\rm f}_i$.

\begin{itemize}[left=5pt]
\item $\tilde{p}_f<v_{\kappa}$: By symmetry, we have 
$\tilde{x}^{\rm f}_i=(1-\sum_{\theta\in\Theta}\tilde{x}^{\rm f}_i)/\kappa<1/\kappa$ for each $i\in [\kappa]$.
\item $\tilde{p}_f=v_{\kappa}$: In this case, there exists a bidder $\theta\in \Theta$ who is still active at $p=\tilde{p}_f$.\footnote{This can be shown by the following arguments: 
Suppose that there is a price $p<v_{\kappa}$ where a bidder in $\Theta$ is dropping and 
only bidders in $[\kappa]$ are active. Then, by the assumption of Case 3, it must hold that 
the new bidder is in the clinching set and thus all the remaining good is allocated by Lemma \ref{dropping_sec5} (iii), 
which contradicts $\tilde{p}_f=v_{\kappa}$.}
By symmetry of bidders, for each $i\in [\kappa-1]$, it holds 
$\tilde{x}_i(v_{\kappa}-)=\tilde{x}_\kappa(v_{\kappa}-)=\tilde{x}^{\rm f}_\kappa$ and 
$\tilde{x}^{\rm f}_i-\tilde{x}_i(v_{\kappa}-)=\tilde{x}^{\rm f}_\theta-\tilde{x}_\theta(v_{\kappa}-)$. 
Then, we have 
$\tilde{x}^{\rm f}_\theta+\tilde{x}^{\rm f}_\kappa= \tilde{x}_\theta(v_{\kappa}-)+\tilde{x}^{\rm f}_i-\tilde{x}_i(v_{\kappa}-)+\tilde{x}_i(v_{\kappa}-)\geq \tilde{x}^{\rm f}_i$.
By $\tilde{x}^{\rm f}_\theta+\tilde{x}^{\rm f}_\kappa+\sum_{i\in [\kappa-1]}\tilde{x}^{\rm f}_i=1$ from (\ref{x_pi}) and Lemma \ref{dropping_sec5} (ii), we have $\tilde{x}^{\rm f}_i\leq 1/\kappa$ for each $i\in [\kappa-1]$.

\item $\tilde{p}_f>v_{\kappa}$: There exists a price $p_2\in (\gamma_1, v_\kappa)$ 
at which bidders in $[\kappa]$ enter the clinching set. 
By the same argument as in the case of $\tilde{p}_f=v_{\kappa}$, there must be a bidder $\theta\in \Theta$ who is still active at $p_2$. Since the number of active bidders is more than $\kappa+1$, 
by the definition of the clinching set, we have 
$\kappa \tilde{b}_i(p_2)/p_2\leq\tilde{S}(p_2)$ for each $i\in[\kappa]$.
By symmetry, for each $i\in [\kappa]$, we have 
\[
\tilde{x}^{\rm f}_i\leq \tilde{x}_i(p_2)+\frac{\tilde{b}_i(p_2)}{p_2}
\leq\frac{1-\tilde{S}(p_2)}{\kappa}+\frac{\tilde{S}(p_2)}{\kappa}=\frac{1}{\kappa}.
\]
\end{itemize}

For property (ii), let $p_3$ be the first price where a new bidder clinches the good.
The key observation is that the future payment of the mechanism is more than $\sum_{i\in [\kappa]}\tilde{b}_i(p_3)$.
This follows by the fact that at price $p_3$, it must hold $\tilde{S}(p_3)\geq\sum_{i\in [\kappa]}\tilde{b}_i(p_3)/p_3$ and 
the remaining good is sold at higher price.
From this and (\ref{x_pi}), property (ii) is shown by 
\begin{align*}
{\rm LW}(x^{\rm f})&\leq \kappa \beta=
\sum_{i\in [\kappa]}\{\tilde{b}_i(p_3)+(\beta-\tilde{b}_i(p_3))\}
\leq  \sum_{i\in [\kappa]\cup \Theta}\tilde{\pi}^{\rm f}_i\leq {\rm LW}(\tilde{x}^{\rm f}),
\end{align*}
where the first inequality holds by Lemma \ref{common_LW_pi} and the last inequality holds by Lemma~\ref{dropping_sec5}~(i).
The same holds for revenue. 
\end{proof}

\section{Monotonicity in Asymmetric Settings}
Now we consider the asymmetric scenario where bidders have different budgets. 
In the limited case of $\tilde{p}_f < p_f$, 
we can easily show that the monotonicity of LW and revenue extends naturally, as shown in Appendix B.
In general, handling asymmetric budgets presents various challenges.  

One such challenge is that the key properties from the symmetric setting are no longer valid. 
The first is the symmetry of the remaining budgets, 
which ensures that the clinching set includes all active bidders if non-empty, 
leading to \eqref{p_s_p_f_relation} and Theorem~\ref{psi}.
Another is Lemma~\ref{common_LW_pi}, which allows for the comparison of both LW and revenue.  
Furthermore, the interaction between a new bidder's valuation and budget affects the outcome in various ways.  
For instance, in the symmetric case, 
Lemma~\ref{change_interval} ensures that the value of $\tilde{\kappa}$ is either $\kappa$ or $\kappa + 1$, 
whereas in the asymmetric case, $\tilde{\kappa}$ can take values outside of this range.

Interestingly, we find counterexamples showing that the allocation monotonicity and LW monotonicity observed in the symmetric case are no longer valid. 
Below, we present these examples. In Example~\ref{2}, 
both the amount of good and the payment for a bidder with $i > \kappa$ increases, 
which occurs by $\tilde{\kappa}>\kappa+1$.

\begin{example}
\label{2}
Consider a multi-unit auction with $n=2$.
Bidder~1 has a valuation of 5 and a budget of 3/2, while bidder~2 has a valuation of 3 
and a budget of 1/2.
In Algorithm \ref{alg:algorithm}, at $p=1/2$, bidder~1 begins clinching the good by $b_2(p)/p=1$. 
As the price increases, only bidder~1 continues to clinch the good 
until either bidder~2 drops out or the remaining budget of bidder 1 reaches 1/2. 
During this period, it follows that $\partial_p b_1(p) = -S(p) = -b_2(p)/p = -1/(2p)$. 
In this case, at $p\in [1/2, 3)$, only bidder~1 clinches the good, which can be seen~by 
\[
3/2-\int^{3}_{1/2}1/(2p)\ {\rm d}p=3/2-1/2\ {\rm log}\, 6 > 1/2.
\]
At $p=3$, bidder 2 drops out of the auction.
Consequently, all the good is allocated to bidder 1 
and their payment is less than their budget due to the above inequality. 
This leads to $\kappa=1$.

Now consider the scenario where bidder 3 is added, with valuation 2 and budget 3. 
For $p \in [0, 2)$, no bidder clinches the good, 
as the condition $\sum_{j \in [3] \setminus i} \tilde{b}_j(p) / p \geq 1$ holds for each $i \in [3]$. 
At $p = 2$, bidder~3 drops out of the auction, and bidder~1 clinches $3/4$ unit while bidder~2 clinches $1/4$ unit, with both payments are equal to their remaining budgets.
Thus, $\tilde{\kappa} = 3$ holds, and the allocation of 
bidder~$\kappa + 1$ might increase in the asymmetric setting.
\end{example}

Example~\ref{1} shows that (i) the amount of good allocated to $i<\kappa$ may increase, 
and (ii) the LW monotonicity does not~hold.

\begin{example}
\label{1}
Consider a multi-unit auction with $n=2$.
Bidder~1 has a valuation of 2 and a budget of 1/3, 
and bidder 2 has a valuation of 1 and a budget of 2/3.
In Algorithm \ref{alg:algorithm}, at $p=1/3$, bidder 2 begins to clinch the good by $b_1(p)/p=1$.
Let $p'$ be the price where the remaining budget of bidder 2 reaches 1/3. 
At $p\in [1/3, p')$, only bidder~2 clinches the good as the price increases 
and it holds $\partial_{p}b_2(p)=-S(p)=-b_1(p)/p=-1/(3p)$ during this period.
Since $p'$ satisfies 
\[
1/3=2/3-\int^{p'}_{1/3}1/(3p){\rm d}p=2/3-1/3\ {\rm log}\, (3p'),
\]
we have $p'={\rm e}/3\simeq 0.906$.
Then, at $p\in [p', 1)$, bidders 1 and 2 have equal remaining budgets and 
clinch the good as the price increases. 
The remaining budget $b_i(p)$ of $i\in [2]$ is changed according to 
$\partial_p b_i(p)=- S(p)=- b_i(p)/p$.
Combining this with $b_i(p')=1/3$, 
we have $b_i(p)=p'/(3p)\ (p'\leq p<1)$. 
By Definitions \ref{derivative}, it holds $\partial_p \Psi_1(p)=-b_i(p)/p^2=-p'/(3 p^3)$ 
and thus we have
\begin{align*}
\Psi_1(p)=1/(3p')-\int^p_{p'} p'/(3 q^3) {\rm d}q= 1/(6p')+p'/(6p^2)
\end{align*}
At $p=1$, bidder 2 drops out of the auction and the remaining good is allocated to bidder 1.
By Theorem \ref{properties_sec5}, we have 
\begin{align*}
x^{\rm f}_1=\Psi_1(1)=1/(6p')+p'/6,\  {\rm and}\ x^{\rm f}_2&=1-x^{\rm f}_1=1-1/(6p')-p'/6.
\end{align*}
Then, ${\rm LW}(x^{\rm f})=\min(v_1 x^{\rm f}_1, 1/3)+\min(v_2 x^{\rm f}_2, 2/3)=1/3+x^{\rm f}_2$.

Consider the case where bidder~3 is added, 
with valuation 2/3 and budget 1. 
At $p=2/3$, bidder~3 is dropping and bidder 2 clinches 0.5 units. 
The remaining budgets of bidder 1 and 2 are both 1/3.
Then, at $p\in [2/3, 1)$, bidders 1 and 2 have equal remaining budgets and 
clinch the good as the price increases. 
Their remaining budget $\tilde{b}_i(p)\ (i\in [2])$ is changed according to 
$\partial_p \tilde{b}_i(p)=- \tilde{S}(p)=- \tilde{b}_i(p)/p$.
Combining this with $\tilde{b}_i(2/3)=1/3$, 
we have $\tilde{b}_i(p)=2/(9p)\ (2/3\leq p<1)$. 
By Definitions \ref{derivative}, it holds $\partial_p \tilde{\Psi}_1(p)=-\tilde{b}_i(p)/p^2=-2/(9 p^3)$. 
By $\tilde{\Psi}_1(2/3)=1/2$, we have
\begin{align*}
\tilde{\Psi}_1(p)=1/2-\int^p_{2/3} 2/(9 q^3) {\rm d}q=1/4+1/(9 p^2)
\end{align*}
Similarly, at $p=1$, bidder 2 drops out of the auction and the remaining good is allocated to bidder~1.
By Theorem \ref{properties_sec5}, we have 
\begin{align*}
\tilde{x}^{\rm f}_1=\Psi_1(1)=13/36\ \ {\rm and}\ \ \tilde{x}^{\rm f}_2&=1-x^{\rm f}_1=23/36\, (<x^{\rm f}_2).
\end{align*}
Then, we have ${\rm LW}(\tilde{x}^{\rm f})=\min(v_1 \tilde{x}^{\rm f}_1, 1/3)+\min(v_2 \tilde{x}^{\rm f}_2, 2/3)
=1/3+\tilde{x}^{\rm f}_2<1/3+x^{\rm f}_2={\rm LW}(x^{\rm f})$.
Thus, the amount of good allocated to bidder $i<\kappa$ is increased and 
LW monotonicity does not hold.
\end{example}

Despite not reaching a definitive conclusion on revenue monotonicity, 
the available evidence suggests that it is still an open question. 
If this property holds, proving it would be a difficult task, perhaps even 
more challenging than the seminal works presented in \citet{BCMX2010} and \citet{GMP2020}.
At the very least, new theoretical tools would be required to compare 
the outcomes before and after the input change.
Instead, in Appendix A, we show how LW and revenue monotonicity 
fail to hold in the indivisible case of adaptive clinching auctions.
Based on the foregoing considerations, it is suggested that in asymmetric settings, when dealing with the addition of online bidders, as in social networks, it might be necessitate 
considerably weaker notions of efficiency and revenue.

\section{Conclusion}
We investigated the monotonicity of adaptive clinching auctions, 
focusing on the change in LW and revenue with an increase in the number of bidders. 
We first examined the symmetric setting, where bidders have equal budgets, and found that 
both LW and revenue are weakly increasing. The results were then extended to include cases 
with new bidders arriving online. Further, we discussed the difficulties in asymmetric settings and provided counterexamples of monotonicity. These findings support the theory of budget-constrained auctions in social networks, where bidders participate through information diffusion. Independently, we developed a useful formula for adaptive clinching auctions with symmetric bidders.

A promising direction for future research involves the investigation of the efficiency goals that can be achieved 
in budget-constrained auctions in social networks. It would be useful to know to what extent properties stronger than non-wastefulness can hold up in clinching auctions, especially in asymmetric settings. 
Moreover, no theoretical guarantees of revenue have been established in asymmetric settings. 
It would also be interesting to investigate the possibility of other mechanisms besides clinching auctions 
that could provide some theoretical guarantees for efficiency and revenue.
	
\paragraph{Acknowledgement}
An extended abstract of this paper will appear in the proceedings of the 24th International Conference on Autonomous Agents and Multiagent Systems (AAMAS'25).
We sincerely thank the anonymous reviewers for their helpful feedback.
This work was supported by Grant-in-Aid for JSPS Research Fellow Grant Number JP22KJ1137, Grant-in-Aid for Challenging Research (Exploratory) Grant Number JP21K19759, and JST ERATO Grant Number JPMJER2301.

\bibliography{monotonicity}

\begin{thebibliography}{33}
\providecommand{\natexlab}[1]{#1}
\providecommand{\url}[1]{\texttt{#1}}
\expandafter\ifx\csname urlstyle\endcsname\relax
  \providecommand{\doi}[1]{doi: #1}\else
  \providecommand{\doi}{doi: \begingroup \urlstyle{rm}\Url}\fi

\bibitem[Ausubel(2004)]{A2004}
L.~M. Ausubel.
\newblock An efficient ascending-bid auction for multiple objects.
\newblock \emph{American Economic Review}, 94\penalty0 (5):\penalty0
  1452--1475, 2004.

\bibitem[Ausubel and Milgrom(2002)]{AM2002}
L.~M. Ausubel and P.~Milgrom.
\newblock Ascending auctions with package bidding.
\newblock \emph{Frontiers of Theoretical Economics}, 1\penalty0 (1), 2002.

\bibitem[Bhattacharya et~al.(2010)Bhattacharya, Conitzer, Munagala, and
  Xia]{BCMX2010}
S.~Bhattacharya, V.~Conitzer, K.~Munagala, and L.~Xia.
\newblock Incentive compatible budget elicitation in multi-unit auctions.
\newblock In \emph{Proceedings of the 21st Annual ACM-SIAM Symposium on
  Discrete Algorithms}, SODA'10, pages 554--572, 2010.

\bibitem[Clarke(1971)]{C1971}
E.~H. Clarke.
\newblock Multipart pricing of public goods.
\newblock \emph{Public Choice}, 11\penalty0 (1):\penalty0 17--33, 1971.

\bibitem[Devanur et~al.(2013)Devanur, Ha, and Hartline]{DHH2013}
N.~R. Devanur, B.~Q. Ha, and J.~D. Hartline.
\newblock Prior-free auctions for budgeted agents.
\newblock In \emph{Proceedings of the 14th ACM Conference on Electronic
  Commerce}, EC'13, pages 287--304, 2013.

\bibitem[Dobzinski and Leme(2014)]{DL2014}
S.~Dobzinski and R.~P. Leme.
\newblock Efficiency guarantees in auctions with budgets.
\newblock In \emph{Proceedings of the 41 st International Colloquium Automata,
  Languages, and Programming}, ICALP'14, pages 392--404, 2014.

\bibitem[Dobzinski et~al.(2012)Dobzinski, Lavi, and Nisan]{DLN2012}
S.~Dobzinski, R.~Lavi, and N.~Nisan.
\newblock Multi-unit auctions with budget limits.
\newblock \emph{Games and Economic Behavior}, 74\penalty0 (2):\penalty0
  486--503, 2012.

\bibitem[Dughmi et~al.(2012)Dughmi, Roughgarden, and Sundararajan]{DRS2012}
S.~Dughmi, T.~Roughgarden, and M.~Sundararajan.
\newblock Revenue submodularity.
\newblock \emph{Theory of Computing}, 8\penalty0 (5):\penalty0 95--119, 2012.

\bibitem[Feng and Hartline(2018)]{FH2018}
Y.~Feng and J.~D. Hartline.
\newblock An end-to-end argument in mechanism design (prior-independent
  auctions for budgeted agents).
\newblock In \emph{Proceedings of the 59th Annual IEEE Symposium on Foundations
  of Computer Science}, FOCS'18, pages 404--415, 2018.

\bibitem[Fikioris and Tardos(2023)]{FT2023}
G.~Fikioris and E.~Tardos.
\newblock Liquid welfare guarantees for no-regret learning in sequential
  budgeted auctions.
\newblock In \emph{Proceedings of the 24th ACM Conference on Economics and
  Computation}, EC'23, pages 678--698, 2023.

\bibitem[Fotakis et~al.(2019)Fotakis, Lotidis, and Podimata]{FLP2019}
D.~Fotakis, K.~Lotidis, and C.~Podimata.
\newblock A bridge between liquid and social welfare in combinatorial auctions
  with submodular bidders.
\newblock In \emph{Proceedings of the 33rd AAAI Conference on Artificial
  Intelligence}, AAAI'19, pages 1949--1956, 2019.

\bibitem[Goel et~al.(2020)Goel, Mirrokni, and Leme]{GMP2020}
G.~Goel, V.~Mirrokni, and R.~P. Leme.
\newblock Clinching auctions with online supply.
\newblock \emph{Games and Economic Behavior}, 123:\penalty0 342--358, 2020.

\bibitem[Gonen and Lerner(2019)]{GL2019}
R.~Gonen and A.~Lerner.
\newblock Towards characterizing the deterministic combinatorial constrained
  efficient space.
\newblock In \emph{Proceedings of the 6th Conference on Algorithmic Decision
  Theory}, ADT'19, pages 32--48, 2019.

\bibitem[Groves(1973)]{G1973}
T.~Groves.
\newblock Incentives in teams.
\newblock \emph{Econometrica}, 41\penalty0 (4):\penalty0 617--631, 1973.

\bibitem[Hirai and Sato(2025)]{HS2025}
H.~Hirai and R.~Sato.
\newblock Polyhedral clinching auctions for indivisible goods.
\newblock \emph{ACM Transactions on Economics and Computation}, 13\penalty0
  (1):\penalty0 4:1--4:30, 2025.

\bibitem[Kawasaki et~al.(2020)Kawasaki, Barrot, Takanashi, Todo, and
  Yokoo]{KBTTY2020}
T.~Kawasaki, N.~Barrot, S.~Takanashi, T.~Todo, and M.~Yokoo.
\newblock Strategy-proof and non-wasteful multi-unit auction via social
  network.
\newblock In \emph{Proceedings of the 34th AAAI Conference on Artificial
  Intelligence}, AAAI'20, pages 2062--2069, 2020.

\bibitem[Li et~al.(2017)Li, Hao, Zhao, and Zhou]{LHZZ2017}
B.~Li, D.~Hao, D.~Zhao, and T.~Zhou.
\newblock Mechanism design in social networks.
\newblock In \emph{Proceedings of the 31st AAAI Conference on Artificial
  Intelligence}, AAAI'17, pages 586--592, 2017.

\bibitem[Li et~al.(2022)Li, Hao, Gao, and Zhao]{LHGZ2022}
B.~Li, D.~Hao, H.~Gao, and D.~Zhao.
\newblock Diffusion auction design.
\newblock \emph{Artificial Intelligence}, 303:\penalty0 103631, 2022.

\bibitem[Li et~al.(2024)Li, Cao, and Zhao]{LCZ2024}
M.~Li, Y.~Cao, and D.~Zhao.
\newblock Double auction on diffusion network.
\newblock In \emph{Proceedings of the 38th AAAI Conference on Artificial
  Intelligence}, AAAI'24, pages 9848--9855, 2024.

\bibitem[Liu et~al.(2023)Liu, Lian, and Zhao]{LLZ2023}
H.~Liu, X.~Lian, and D.~Zhao.
\newblock Diffusion multi-unit auctions with diminishing marginal utility
  buyers.
\newblock In \emph{Proceedings of the 22nd International Conference on
  Autonomous Agents and Multiagent Systems}, AAMAS'23, pages 2715--2717, 2023.

\bibitem[Lu and Xiao(2015)]{LX2015}
P.~Lu and T.~Xiao.
\newblock Improved efficiency guarantees in auctions with budgets.
\newblock In \emph{Proceedings of the 16th ACM Conference on Economics and
  Computation}, EC'15, pages 397--413, 2015.

\bibitem[Lu and Xiao(2017)]{LX2017}
P.~Lu and T.~Xiao.
\newblock Liquid welfare maximization in auctions with multiple items.
\newblock In \emph{Proceedings of the 10th International Symposium on
  Algorithmic Game Theory}, SAGT'17, pages 41--52, 2017.

\bibitem[Milgrom(2004)]{M2004}
P.~Milgrom.
\newblock \emph{Putting Auction Theory to Work}.
\newblock Cambridge University Press, Cambridge, 2004.

\bibitem[Parkes(2007)]{P2007}
D.~C. Parkes.
\newblock Online mechanisms.
\newblock In N.~Nisan, T.~Roughgarden, E.~Tardos, and V.~V. Vazirani, editors,
  \emph{Algorithmic Game Theory}, chapter~16, pages 411--440. Cambridge
  University Press, 2007.

\bibitem[Rastegari et~al.(2011)Rastegari, Condon, and Leyton-Brown]{RCL2011}
B.~Rastegari, A.~Condon, and K.~Leyton-Brown.
\newblock Revenue monotonicity in deterministic, dominant-strategy
  combinatorial auctions.
\newblock \emph{Artificial Intelligence}, 175\penalty0 (2):\penalty0 441--456,
  2011.

\bibitem[Sato(2023)]{S2023}
R.~Sato.
\newblock Polyhedral clinching auctions with a single sample.
\newblock \emph{arXiv: 2302.03458}, 2023.

\bibitem[Syrgkanis and Tardos(2013)]{ST2013}
V.~Syrgkanis and E.~Tardos.
\newblock Composable and efficient mechanisms.
\newblock In \emph{Proceedings of the 45th Annual ACM Symposium on Theory of
  Computing}, STOC'13, pages 211--220, 2013.

\bibitem[Takanashi et~al.(2019)Takanashi, Kawasaki, Todo, and Yokoo]{TKTY2019}
S.~Takanashi, T.~Kawasaki, T.~Todo, and M.~Yokoo.
\newblock Efficiency in truthful auctions via a social network.
\newblock \emph{arXiv: 1904.12422}, 2019.

\bibitem[Vickrey(1961)]{V1961}
W.~Vickrey.
\newblock Counterspeculation, auctions, and competitive sealed tenders.
\newblock \emph{The Journal of Finance}, 16\penalty0 (1):\penalty0 8--37, 1961.

\bibitem[Xiao et~al.(2022)Xiao, Song, and Khoussainov]{XSK2022}
M.~Xiao, Y.~Song, and B.~Khoussainov.
\newblock Multi-unit auction in social networks with budgets.
\newblock In \emph{Proceedings of the 36th AAAI Conference on Artificial
  Intelligence}, AAAI'22, pages 5228--5235, 2022.

\bibitem[Xu et~al.(2019)Xu, He, and Zhao]{XHZ2019}
J.~Xu, X.~He, and D.~Zhao.
\newblock Double auction design on networks.
\newblock In \emph{Proceedings of the First International Conference on
  Distributed Artificial Intelligence}, DAI '19, pages 1--6, 2019.

\bibitem[Zhao(2022)]{Z2022}
D.~Zhao.
\newblock Mechanism design powered by social interactions: A call to arms.
\newblock In \emph{Proceedings of the 31st International Joint Conference on
  Artificial Intelligence}, IJCAI'22, pages 5831--5835, 2022.

\bibitem[Zhao et~al.(2018)Zhao, Li, Xu, Hao, and Jennings]{ZLXHJ2018}
D.~Zhao, B.~Li, J.~Xu, D.~Hao, and N.~R. Jennings.
\newblock Selling multiple items via social networks.
\newblock In \emph{Proceedings of the 17th International Conference on
  Autonomous Agents and MultiAgent Systems}, AAMAS'18, pages 68--76, 2018.

\end{thebibliography}

\appendix
\section{Indivisible Settings}
	Let $\mathbf Z_{+}$ denote the set of nonnegative integers.
	In this section, we consider the indivisible setting where 
	the seller sells $l\in \mathbf Z_{+}$ units of homogeneous goods and 
	the allocation must be an integer vector, i.e., $x\in \mathbf Z^N_+$.
	We first explain the adaptive clinching auction for indivisible goods proposed by \citet{DLN2012}.
	The mechanism can be described as follows:
	\begin{algorithm}[bth]
	\caption{Adaptive Clinching Auction for Indivisible Goods~\citep{DLN2012}}
	\label{adaptive_clinching} 
	\begin{algorithmic}[1]
	  \STATE $x_i=0,\ \pi_i:=0,\ d_i:=l+1 \ 
	  (i\in N)\ {\rm and}\ c:=0$.\\
	  \WHILE{active bidders exist}
	  \STATE Increase $c$ until there appears an active bidder 
	  $j$ with $v'_{j}=c$ or $d_{j}=(B_j-\pi_j)/c$.
	  \WHILE{$\exists$ active bidder $j$ with $v'_{j}=c$}
	  \STATE Pick such a bidder $j$ and let $d_{j}:=0$.
	  \FOR{$i=1,2,\ldots,n$} 
	  \STATE $\delta_i= \max(l-\sum_{k\neq i}d_k,0), x_i:=x_i+\delta_i,\ 
	  \pi_i:=\pi_i+c \delta_i,\ d_i:=d_i-\delta_i$.
	  \STATE $l=l-\delta_i$.
	  \ENDFOR
	  \ENDWHILE
	  \WHILE{$\exists$ active bidder $j$ with $d_j=(B_j-\pi_j)/c$}
	  \STATE Pick such a bidder $j$ and let $d_{j}:=d_{j}-1$.
	   \FOR{$i=1,2,\ldots,n$} 
	  \STATE $\delta_i= \max(l-\sum_{k\neq i}d_k,0), x_i:=x_i+\delta_i,\ 
	  \pi_i:=\pi_i+c \delta_i,\ d_i:=d_i-\delta_i$.
	  \STATE $l=l-\delta_i$.
	  \ENDFOR
	  \ENDWHILE
	 \ENDWHILE
	 \STATE Output $(x,\pi)$.
	\end{algorithmic}
	\end{algorithm}

	The variables $c,\delta_i, x_i,\pi_i,d_i, l$ are interpreted as follows:
\begin{itemize}
\item $c$ is the price clock. It represents the current transaction price per unit. It starts at zero and gradually increases.
\item $\delta_i$ is the number of the goods clinched by bidder $i$ in that iteration.
\item $x_i$ is the number of the goods allocated to bidder $i$. 
\item $\pi_i$ is the payment of bidder $i$.
\item $d_i$ is the demand of bidder $i$, which represents the number of maximum possible amounts that bidder $i$ can clinch at the current price.
\item $l$ is the number of the remaining good.
\end{itemize}
	
Similar to Algorithm~\ref{alg:algorithm}, the price gradually rises and the demands of the bidders 
decrease accordingly. Also, if the total demand of other bidders is less than the number of the remaining good, the bidder clinches the goods at the current price.  After updating the allocation and the demand, the process is repeated until $d_i=0$ for each $i\in N$. Consequently, the final allocation is output. 
	
The following examples are counterexamples of LW monotonicity and revenue monotonicity
for indivisible goods.
\begin{example}
Consider a multi-unit auction with $n=2$ and four units of the goods.
Bidder 1 has a valuation of 6 and a budget of 9, and bidder 2 has a valuation of 9 and a budget of 15.
In the adaptive clinching auction, bidder 1 clinches one unit of the goods and bidder 2 clinches three units.
Then, the LW is $\min(6,9)+\min(15,15)=21$.

Now consider the case where bidder 3 is added.
The valuation and budget of bidder 3 are 4 and $\infty$, respectively. 
Then, bidder 1 and bidder 2 clinch 2 units each, and bidder~3 does not clinch any goods.
The LW is $\min(12,9)+\min(10,15)=19<21$.
Thus, LW monotonicity does not always~hold.
\end{example}
\begin{example}
Consider a multi-unit auction with $n=2$ and five units of the goods.
We first consider the input $N=3$, $v=[6,5,4]$, and $B=[4, 2, 2]$.
In the adaptive clinching auction, 
bidder 1 clinches one unit of the goods at $p=2/3$,
bidder~1 clinches two units at $p=1$,
and bidder 2 and bidder 3 clinch one unit each at $p=4/3$.
Then, the revenue is $16/3$. 

Now consider the case where $N^\theta=4$, $v^\theta=[6,5,4,1]$, and $B^\theta=[4, 2, 2, \infty]$. 
bidder~1 clinches three units, bidder 2 and bidder 3 clinch one unit each at $p=1$.
Then, the revenue is $5(<16/3)$. 
Thus, revenue monotonicity does not always hold.
\end{example}

\section{Limited Case with Asymmetric Budgets}
We show that, in the limited case where $\tilde{p}_f < p_f$, 
the monotonicity of LW and revenue extends naturally to asymmetric settings. 
Note that all notation has been carried over from Section 4.
\begin{lemma}
\label{limited-general}
In Algorithm~\ref{alg:algorithm}, if $\tilde{p}_f<p_f$, 
then it holds ${\rm LW}(\tilde{x}^{\rm f})\geq{\rm LW}(x^{\rm f})$ and 
$ {\rm REV}(\tilde{\pi}^{\rm f})\geq {\rm REV}(\pi^{\rm f})$.
\end{lemma}
\begin{proof}
By Theorem \ref{properties_sec5}, it holds $v_i \tilde{x}^{\rm f}_i\geq \tilde{\pi}^{\rm f}_{i}=B_i$ for each $i\in N^\theta$ with $v_i>\tilde{p}_f$.
By $v_\kappa=p_f>\tilde{p}_f$, such bidders include all bidders in~$[\kappa]$.
Then, we have
\begin{align*}
{\rm LW}(\tilde{x}^{\rm f})
\geq\sum_{i\in N^\theta; v_i\geq \tilde{p}_f}B_i
\geq\sum_{i\in [\kappa]}B_i
\geq {\rm LW}(x^{\rm f}),
\end{align*}
where the last equality holds by $x^{\rm f}_i=0$ for $i>\kappa$ from Theorem \ref{properties_sec5} (iii). 
The same is true if we replace LW with revenue.
\end{proof}

The proof idea is close to Theorem 4.5 of \citet{DL2014}, which shows that the LW of their unit-price auction is more than that of the adaptive clinching auction. Their mechanism is to calculate the market clearing price and sell all the good at the price. It is noteworthy that the execution of their mechanism can be reproduced by the adaptive clinching auction by adding a bidder with a valuation equal to the market clearing price and an infinite budget. 
Furthermore, in many of such cases, it holds $\tilde{p}_f<p_f$.
For example, consider the market in Example 3.7, and suppose that a new bidder with a valuation of 2 and an infinite budget is added. Then, no bidder clinches the good until the price reaches 2, the new bidder drops out, and bidders 1 and 2 clinch 1/2 unit each. This is an outcome identical to the unit-price auction applied to bidders 1 and 2. 

\section{Omitted Proofs}
\subsection{Proof of Lemma \ref{p_s_p_f_common}}
Suppose that $\kappa=1$. Then, by Theorem \ref{properties_sec5} (iii), the claim trivially holds by~$\max(v_2, 0)=p_s<\beta$.
Suppose that $\kappa\geq 2$.
Firstly, we show $p_s<\kappa \beta$.
By Theorem \ref{properties_sec5} (ii) and (iii), it holds $1=\sum_{i\in N}x^{\rm f}_i=\sum_{i=1}^{\kappa-1}x^{\rm f}_i+x^{\rm f}_{\kappa}$. 
Since $p_s$ is the lowest price where bidders clinch the good, we have $x^{\rm f}_i\leq \pi^{\rm f}_i/p_s$ for each $i\in [\kappa]$. 
Then, we have 
$1=\sum_{i=1}^{\kappa-1}x^{\rm f}_i+x^{\rm f}_{\kappa}\leq  (\kappa-1)\beta/p_s+\pi^{\rm f}_{\kappa}/p_s< \kappa \beta/p_s$,
where the last inequality holds by $\pi^{\rm f}_\kappa<\beta$ (by the definition of $\kappa$). 
Therefore, we have $p_s<\kappa \beta$. 
Secondly, we show $\max(v_{\kappa+1}, (\kappa-1) \beta)= p_s$.
Suppose that $p<\min(v_{\kappa+1}, p_s)$, i.e., $p<\min(v_{\kappa+1}, p_s)\leq p_s<\kappa \beta$. Then, by $p<p_s$, the remaining budgets of active bidders are~$\beta$.
Since the number of active bidders is more than $\kappa+1$, at the moment, it holds $\sum_{j\in A(p)\setminus i}\beta/p\geq\kappa \beta/p>1=S(p)$ for each $i\in A(p)$.
This means $C(p)=\emptyset$ and $|C(p_s)|<\kappa+1$.
Thus, it holds $p_s\geq v_{\kappa+1}$ and no bidder clinches the good 
when~$p<\min(v_{\kappa+1}, p_s)=v_{\kappa+1}$.

To show $p_s=\max(v_{\kappa+1}, (\kappa-1) \beta)$, 
we perform the following case-by-case~analysis:

Case 1: $(\kappa-1)\beta\leq v_{\kappa+1}$. At $p=v_{\kappa+1}$, the number of active bidder is decreased to~$\kappa$.
Then, in line 6 of the iteration, it holds 
\[
\sum_{j\in A(p)\setminus i}b_i(p-)/p =(\kappa-1) \beta/v_{\kappa+1}\leq 1=S(p)\ \ (i\in A(p)).
\] 
This implies $p_s=v_{\kappa+1}=\max(v_{\kappa+1}, (\kappa-1) \beta)$.

Case 2: $v_{\kappa+1}< (\kappa-1)\beta\leq v_{\kappa}$.
At $p=v_{\kappa+1}$, the number of active bidder is decreased to $\kappa$.
Then, in line 6 of the iteration, it~holds 
\[
\sum_{j\in A(p)\setminus i}b_i(p-)/p=(\kappa-1) \beta/v_{\kappa+1}>1=S(p)\ \ (i\in A(p)),
\]
which implies $v_{\kappa+1}< p_s\leq (\kappa-1)\beta$.
At $p=(\kappa-1)\beta\leq v_{\kappa}$, the number of active bidders is at least $\kappa-1$.
Then, it holds 
\[
\sum_{j\in A(p)\setminus i}b_i(p-)/p\leq (\kappa-1)\beta/(\kappa-1)\beta=1=S(p),
\] 
which means $p_s= (\kappa-1)\beta=\max(v_{\kappa+1}, (\kappa-1) \beta)$.

Case 3: $v_{\kappa+1}< v_{\kappa}< (\kappa-1)\beta$. 
Since all the good are traded at the prices less than $p_f(=v_{\kappa})$, it holds 
$x^{\rm f}_i\geq \pi^{\rm f}_i/p_f=\beta/ v_{\kappa}>1/(\kappa-1)$ for each $i\in [\kappa-1]$, 
where the equality holds by $ \pi^{\rm f}_i=\beta$ and $p_f=v_\kappa$ (Theorem~\ref{properties_sec5}~(iii)).
This means 
\[
\sum_{i\in N}x^{\rm f}_i\geq \sum_{i\in [\kappa-1]}x^{\rm f}_i>(\kappa-1)/(\kappa-1)=1,
\]
which contradicts with Theorem~\ref{properties_sec5}~(ii). 
Thus, this case never happens (and we have $(\kappa-1)\beta\leq v_{\kappa}$).

By the above arguments, we have $\max(v_{\kappa+1}, (\kappa-1) \beta)=p_s<~\kappa\beta$.

\subsection{Proof of Lemma \ref{change_interval}}
We perform the case-by-case~analysis. 
By $v_\theta\neq v_\kappa$ and $p_s<\kappa \beta$ (Lemma~\ref{p_s_p_f_common}), the following cases contain all the possible scenarios.

Case 1: $v_{\theta}\leq p_s$. In this case, bidder $\theta$ drops out of the auction before bidders begin to clinch the good. Then, the allocation is unchanged. 
Thus, it holds $(\tilde{p}_s, \tilde{p}_f,\tilde{\kappa})=(p_s, p_f, \kappa)$.

Case 2: $\kappa=1$ and $p_s<\min(v_{\theta}, v_1)< \beta$. 
Since $\min(v_{\theta}, v_1)$ is the second highest valuation, 
by Theorem \ref{properties_sec5} (iii) and $\min(v_{\theta}, v_1)< \beta$, we have $\tilde{\kappa}=1$.
Moreover, again by Theorem \ref{properties_sec5} (iii), we also have $\tilde{p}_s=\tilde{p}_f=\min(v_{\theta}, v_1)$.
Therefore, we have $(\tilde{p}_s, \tilde{p}_f,\tilde{\kappa})=(\min(v_{\theta}, v_1), \min(v_{\theta}, v_1), 1)$.

In the remaining cases, the second highest valuation in $v^\theta$ is not less than $\beta$.
In Cases~3 and 4, by $\kappa\geq 2$ and Lemma \ref{p_s_p_f_common}, we consider the case of $\beta\leq p_s\leq\min(v_\theta,v_\kappa)$. 
In Cases~5 and 6, we consider the case of $\beta\leq \kappa\beta\leq\min(v_\theta,v_\kappa)$. 
Therefore, in these cases, it must be $\tilde{\kappa}\geq 2$ by Theorem \ref{properties_sec5} (iii).

We first show $\tilde{p}_s\geq \min(v_{\theta}, v_\kappa, \kappa \beta)$ in these cases.
By Lemma~\ref{p_s_p_f_common}, we have $v_{\kappa+1}\leq p_s\leq \min(v_{\theta}, v_{\kappa}, \kappa \beta)$.
At $p<\min(v_{\theta}, v_\kappa, \kappa \beta)$, it holds $|\tilde{A}(p)|\geq \kappa+1$ and 
$\sum_{j\in \tilde{A}(p)\setminus i}\beta/p\geq\kappa\beta/p>\kappa \beta/\kappa \beta=1\geq \tilde{S}(p)$ for each $i\in \tilde{A}(p)$.
Note that the second inequality in the above holds by 
$p<\min(v_{\theta}, v_\kappa, \kappa \beta)\leq \kappa \beta$. 
This implies $\tilde{C}(p)=\emptyset$ and $\tilde{p}_s\geq \min(v_{\theta}, v_\kappa, \kappa \beta)$. 
In the following, we show that $\tilde{p}_s=\min(v_{\theta}, v_\kappa, \kappa \beta)$ in all the remaining cases.

Case 3: $\kappa\geq 2$ and $p_s< v_{\kappa}<\min(v_{\theta}, \kappa \beta)$. 
When $p=v_{\kappa}$, the number of active bidder is decreased to $\kappa$. 
In line 6 of the iteration, it holds \[
\tilde{\delta}_i=\max\bigl(0,\tilde{S}(p-)-\sum_{j\in \tilde{A}(p)\setminus i}b_j(p-)/p\bigr)=\max\bigl(0,1-(\kappa-1) \beta/v_{\kappa}\bigr)
\] 
for each $i\in \tilde{A}(v_\kappa)$.
Since it holds $(\kappa-1)\beta\leq p_s< v_{\kappa}<\kappa \beta$ (Lemma \ref{p_s_p_f_common}),
we have $0< \tilde{\delta}_i<1-(\kappa-1)/\kappa=1/\kappa$.
By $|\tilde{A}(v_\kappa)|=\kappa$, this means $0<\sum_{i\in \tilde{A}(v_{\kappa})}\tilde{\delta}_i<1$ 
and we have $\tilde{p}_s=v_{\kappa}$.
Then, by (\ref{p_s_p_f_relation}) and $\tilde{\kappa}\geq 2$, we have $\tilde{p}_f=\min(v_{\theta},v_{\kappa-1})$.
Since this bidder has the $\kappa$-th highest valuation in $v^{\theta}$, 
we also have $\tilde{\kappa}=\kappa$ by (\ref{p_s_p_f_relation}).
Thus, we have $(\tilde{p}_s, \tilde{p}_f,\tilde{\kappa})=(v_{\kappa}, \min(v_{\theta},v_{\kappa-1}), \kappa)$.

Case 4: $\kappa\geq 2$ and $p_s< v_{\theta}<\min(v_{\kappa}, \kappa  \beta)$. 
At $p=v_{\theta}$,  the number of active bidder is decreased to $\kappa$. 
In line 6 of the iteration, it holds 
\[
\tilde{\delta}_i=\max\bigl(0,\tilde{S}(p-)-\sum_{j\in \tilde{A}(p)\setminus i}b_j(p-)/p\bigr)=\max\bigl(0,1-(\kappa-1) \beta/v_{\theta}\bigr)
\] 
for each $i\in \tilde{A}(v_\theta)$.
Since it holds $(\kappa-1)\beta\leq p_s< v_{\theta}<\kappa \beta$ (Lemma~\ref{p_s_p_f_common}),
we have $0< \tilde{\delta}_i<1-(\kappa-1)/\kappa=1/\kappa$.
By $|\tilde{A}(v_\theta)|=\kappa$, this means $0<\sum_{i\in \tilde{A}(v_{\theta})}\tilde{\delta}_i<1$.
Thus, we have $v_\theta=\tilde{p}_s<\tilde{p}_f$.
Then, by (\ref{p_s_p_f_relation}) and $\tilde{\kappa}\geq 2$, we have $\tilde{p}_f=v_{\kappa}$.
Since bidder $\kappa$ has the $\kappa$-th highest valuation in $v^{\theta}$, we have $\tilde{\kappa}=\kappa$.
Thus, we have $(\tilde{p}_s, \tilde{p}_f,\tilde{\kappa})=(v_{\theta}, v_{\kappa},\kappa)$.

Case 5: $p_s<\kappa \beta<\min(v_{\theta}, v_{\kappa})$. 
Combining this with $v_{\kappa+1}\leq p_s$ (Lemma  \ref{p_s_p_f_common}), it holds $\tilde{A}(\kappa\beta)=\kappa+1$.
At $p=\kappa\beta$, it holds 
$\sum_{j\in \tilde{A}(p)\setminus i}\tilde{b}_j(p-)/p=\kappa\times \beta/\kappa \beta=1=\tilde{S}(p)$ for each $i\in \tilde{A}(p)$.
Therefore, we have $\tilde{p}_s=\kappa \beta$ and $\tilde{C}(p)=\tilde{A}(p)$.
Combining~(\ref{p_s_p_f_relation}) and $\tilde{\kappa}\geq 2$ with $v_{\kappa+1}\leq p_s<\tilde{p}_s<\min(v_{\theta}, v_{\kappa})$,
we have $\tilde{p}_f=\min(v_{\theta}, v_{\kappa})$.
Since this bidder is the $(\kappa+1)$-th highest valuation in $v^{\theta}$, we have $\tilde{\kappa}=\kappa+1$ by (\ref{p_s_p_f_relation}).
Therefore, we have $(\tilde{p}_s, \tilde{p}_f,\tilde{\kappa})=(\kappa \beta, \min(v_{\theta}, v_{\kappa}), \kappa+1)$.

Case 6: $p_s< \kappa \beta=\min(v_{\theta}, v_{\kappa})$. 
Then, by Lemma~\ref{p_s_p_f_common}, it holds 
$(\kappa-1)\beta\leq p_s< \kappa \beta$.
At $p=\kappa \beta=\min(v_{\theta}, v_{\kappa})$, the number of active bidder is decreased to $\kappa$. 
For each $i\in \tilde{A}(p)$, it holds 
\[
\tilde{\delta}_i=\max\bigl(0,\tilde{S}(p-)-\sum_{j\in \tilde{A}(p)\setminus i}\tilde{b}_j(p-)/p\bigr)=1-(\kappa-1)\beta/\kappa\beta=1/\kappa
\] 
in line 6 of this iteration.
Then, all the good are allocated to $\kappa$ bidders in this iteration, and thus we have $\tilde{p}_s=\tilde{p}_f=\kappa \beta$.
The payment of these $\kappa$ bidders are $\kappa \beta\times 1/\kappa=\beta$, which means $\tilde{\kappa}=\kappa+1$.
Therefore, we have $(\tilde{p}_s, \tilde{p}_f,\tilde{\kappa})=(\kappa \beta, \kappa \beta, \kappa+1)$.

\subsection{Proof of Lemma \ref{dropping_sec5}}
(i) By budget-feasibility and IR, it holds $v_i \tilde{x}^{\rm f}_i\geq \tilde{\pi}^{\rm f}_i$ and $\beta\geq \tilde{\pi}^{\rm f}_i$ for each $i\in N$. Thus, we have $\min(v_i \tilde{x}^{\rm f}_i, \beta)\geq \tilde{\pi}^{\rm f}_i$ for each $i\in N^\Theta$.

(ii) (iii) Suppose that $i$ is the first bidder who drops out of the auction among the bidders in the clinching set. Since $i$ is in the clinching set, just before the price $p$ reaches $v_i$, it holds 
$\tilde{S}(v_i-)=\sum_{j\in  \tilde{A}(v_i)}\tilde{b}_j(v_i-)/v_i$.
Then, in line 6 of the iteration, for each $j\in  \tilde{A}(v_i)$, it holds 
\[
\tilde{\delta}_j=\tilde{S}(v_i-)-\sum_{k\in  \tilde{A}(v_i)\setminus j}\tilde{b}_k(v_i-)/v_i=\tilde{b}_j(v_i-)/v_i.
\]
Moreover, in line 7 of the iteration, for each $j\in  \tilde{A}(v_i)$, it holds 
$\tilde{x}_j(v_i)=\tilde{x}_j(v_i-)+\tilde{\delta}_j=\tilde{x}_j(v_i-)+\tilde{b}_j(v_i-)/v_i=\tilde{\Psi}_j(v_i-)=\tilde{\Psi}_j(v_i)$ 
and $\tilde{b}_j(v_i)=\tilde{b}_j(v_i-)-v_i\tilde{\delta}_j=0$.
In this iteration, each bidder $j\in  \tilde{A}(v_i)$ clinches $\tilde{b}_j(v_i-)/v_i$ units. 
In total, the number of the goods allocated in the iteration is 
\[
\sum_{j\in  \tilde{A}(v_i)}\tilde{\delta}_j=\sum_{j\in  \tilde{A}(v_i)}\tilde{b}_j(v_i-)/v_i=\tilde{S}(v_i-),
\]
which means that all the remaining good are allocated to the remaining active bidders.
Therefore, it holds $\tilde{p}_f=v_i$ and 
all the good are sold at the end of the auction, i.e., $\sum_{i\in N^\Theta}\tilde{x}^{\rm f}_i=1$.
Furthermore, by the above arguments, it holds $\tilde{x}^{\rm f}_i=\tilde{\Psi}_i(\tilde{p}_f)$ and $\tilde{\pi}^{\rm f}_i=\beta$ for each $i\in \tilde{A}(\tilde{p}_f)$. Also, it holds $\tilde{x}^{\rm f}_i=\tilde{x}_i(\tilde{p}_f-)$ and $\tilde{\pi}^{\rm f}_i=\beta-\tilde{b}_i(\tilde{p}_f-)$ for $i$ with $v_i=\tilde{p}_f$.

\subsection{Proof of the claim in Case 2-1 of Theorem \ref{tildepsi_relation}}
In Case 2, we only consider scenarios where new bidders do not clinch the good. We can assume without loss of generality that there is at most one active new bidder at each price in $[\gamma_1, v_\kappa)$. Even if there are multiple active new bidders, we can treat these active new bidders as a single bidder who arrives at the earliest time among them and has the highest valuation among them.

Now we consider the following situation: When bidders in $[\kappa]$ are in the clinching set, and a new bidder $\theta$ with a valuation $v_\theta$ arrives at price $p =\gamma$. 
After the arrival of $\theta$, by the assumption of Case 2, 
no bidder is in the clinching set as long as bidder $\theta$ is active\footnote{Otherwise, bidder $\theta$ clinches some good.}. When $\theta$ is dropping at $p = v_\theta$, all bidders in $[\kappa]$ enter the clinching set and clinch some~good. 

Let $p'$ be any price in $[v_\theta, \gamma')$, where $\gamma'$ is the minimum of $\tilde{p}_f$ and the price at which the next new bidder arrives. In the following, we show that the values of $\tilde{b}(p')$ and $\tilde{\Psi}(p')$ are less than those in the scenario where $\theta$ had not participated. 
If we can establish this, the claim is shown by repeatedly applying it for each interval where a new bidder is active,

Let $\alpha$ be the remaining budget of bidders in $[\kappa]$ 
when a new bidder $\theta\in \Theta$ with $v_\theta<v_\kappa$ arrives at $\gamma$.
Remark that, by the symmetry of bidders, it holds 
$\alpha:=\tilde{b}_i(\gamma)$ for each $i\in [\kappa]$.
At $p=\gamma$, since all bidders in $[\kappa]$ are in the clinching set, it holds  
$\tilde{S}(\gamma)=(\kappa-1)\alpha/\gamma$ for each $i\in [\kappa]$.
By $\alpha<\beta$ and the assumption of Case 2, 
no bidder clinches the good before bidder $\theta$ drops out at $p=v_\theta<v_\kappa$.
In line 6 of the iteration at $p=v_\theta$, each bidder $i\in [\kappa]$ enters the clinching set by 
\[
\tilde{S}(v_\theta)=\tilde{S}(\gamma)=(\kappa-1)\alpha/\gamma
>(\kappa-1)\alpha/v_\theta=(\kappa-1)\tilde{b}_i (v_\theta-)/v_\theta.
\]
Also, the amount of the good that bidder $i$ clinches in line 6 is 
\[
\tilde{\delta}_i=\sum_{j\in A(p)\setminus i}\bigl(\frac{\alpha}{\gamma}-\frac{\alpha}{v_\theta}\bigr)
=(\kappa-1)\bigl(\frac{1}{\gamma}-\frac{1}{v_\theta}\bigr)\alpha.
\]
The remaining budget of bidder $i$ at the end of the iteration~is 
\begin{align*}
\tilde{b}_i(v_\theta)
=\bigl\{1-\frac{v_\theta(\kappa-1)}{\gamma}+(\kappa-1)\bigr\}\alpha
=\bigl(\kappa-\frac{v_\theta(\kappa-1)}{\gamma}\bigr)\alpha.
\end{align*}
Before the price reaches $\gamma'$,  
in line 3 of the next iteration, 
$\tilde{S}(p)$ is changed with keeping $\tilde{S}(p)=(\kappa-1)\tilde{b}_i(p)/p$ as in~(\ref{SP}).
The remaining budget $\tilde{b}_i(p)$ of $i\in [\kappa]$ is changed according to  
$\partial_p b_i(p)=- \tilde{S}(p)=-(\kappa-1)\tilde{b}_i(p)/p$.
Since $\tilde{b}_i(p)$ is continuous at the interval $[v_\theta, \gamma')$, solving the differential equation, we have 
\begin{align*}
\tilde{b}_i(p')&=\tilde{b}_i(v_\theta)(\frac{v_\theta}{p'})^{\kappa-1}=
(\frac{v_\theta}{p'})^{\kappa-1}\bigl(\kappa-\frac{v_\theta(\kappa-1)}{\gamma}\bigr)\alpha.
\end{align*}
Since $\tilde{\Psi}_i(p')$ is continuous, by Lemma~\ref{derivative}, we have 
\begin{align*}
\tilde{\Psi}_i(p')=\tilde{\Psi}_i(\gamma)-\frac{\alpha}{\gamma}+\frac{\alpha}{v_\theta}-\int^{p'}_{v_\theta} \frac{\tilde{b}_i(q)}{q^2}{\rm d}q.
\end{align*}
For the third term, we have
\begin{align*}
-\int^{p'}_{v_\theta} \frac{\tilde{b}_i(q)}{q^2}{\rm d}q&=-(\kappa-\frac{v_\theta (\kappa-1)}{\gamma})\alpha\int^{p'}_{v_\theta} (\frac{v_\theta}{q})^{\kappa-1}\frac{1}{q^2}{\rm d}q\\
&=(\kappa-\frac{v_\theta (\kappa-1)}{\gamma})\frac{v_\theta^{\kappa-1}}{\kappa}(\frac{1}{(p')^\kappa}-\frac{1}{v_\theta^\kappa})\alpha\\
&= \bigl(-\frac{1}{v_\theta}+\frac{1}{\gamma}-\frac{1}{\kappa\gamma}+\frac{(\kappa-\frac{v_\theta (\kappa-1)}{\gamma})v_\theta^{\kappa-1}}{\kappa (p')^\kappa} \bigr)\alpha.
\end{align*}
Therefore, we have 
\begin{align*}
\tilde{\Psi}_i(p')&=\tilde{\Psi}_i(\gamma)- \frac{1}{\kappa\gamma}\alpha+\frac{(\kappa-\frac{v_\theta (\kappa-1)}{\gamma})v_\theta^{\kappa-1}}{\kappa (p')^\kappa}\alpha.
\end{align*}

Now we consider the case where bidder $\theta$ does not exist.
Let $\ddot{b}_i(p)$ and $\ddot{\Psi}(p)$ be the corresponding functions in that case.
Since $\ddot{b}_i(p)$ and $\ddot{\Psi}(p)$ are continuous, by the same argument as the proof of Theorem \ref{psi}, 
we have 
\begin{align*}
\ddot{b}_i(p')&=(\frac{\gamma}{p'})^{\kappa-1}\alpha\\
\ddot{\Psi}_i(p')&=\tilde{\Psi}_i(\gamma)-\int^{p'}_{\gamma} \frac{\tilde{b}_i(q)}{q^2}{\rm d}q
=\tilde{\Psi}_i(\gamma)- \frac{1}{\kappa\gamma}\alpha+\frac{\gamma^{\kappa-1}}{\kappa (p')^\kappa}\alpha.
\end{align*}

To compare them with $\tilde{b}_i(p')$ and $\tilde{\Psi}_i(p')$, we now show
$(\frac{v_\theta}{\gamma})^{\kappa-1}(\kappa-\frac{v_\theta(\kappa-1)}{\gamma})\leq 1$.
Define a function $H_{\kappa}(x)$ by $H_{\kappa}(x):=x^{\kappa-1}(\kappa-\kappa x+x)$ for $x>0$.
Then, we first show that $H_{\kappa}(x)$ is weakly decreasing on $x>1$.
If $\kappa=1$, then $H_{\kappa}(x)$ is a constant function by 
$H_{\kappa}(x)=x^{\kappa-1}(\kappa-\kappa x+x)=1-x+x=1$.
If $\kappa\geq 2$, we have 
$H_{\kappa}'(x)=(\kappa-1)\kappa x^{\kappa-2} (1-x)<0$,
where the last inequality holds by $x>1$. Therefore, in both cases,  $H_{\kappa}(x)$ is weakly decreasing on $x>1$.
By $v_\theta/\gamma>1$, we have 
\[
(\frac{v_\theta}{\gamma})^{\kappa-1}(\kappa-\frac{v_\theta(\kappa-1)}{\gamma})=H_{\kappa}(v_\theta/\gamma)\leq H_{\kappa}(1)=\kappa-\kappa+1=1.
\]
Using the inequality, we obtain the following relations:
\begin{align*}
\tilde{\Psi}_i(p')&=\tilde{\Psi}_i(\gamma)- 
\frac{1}{\kappa\gamma}\alpha+(\frac{v_\theta}{\gamma})^{\kappa-1}(\kappa-\frac{v_\theta(\kappa-1)}{\gamma})\frac{\gamma^{\kappa-1}}{\kappa (p')^\kappa}\alpha\\
&\leq \tilde{\Psi}_i(\gamma)- \frac{1}{\kappa\gamma}\alpha+\frac{\gamma^{\kappa-1}}{\kappa (p')^\kappa}\alpha
=\ddot{\Psi}_i(p'),\\
\tilde{b}_i(p')&
=(\frac{v_\theta}{\gamma})^{\kappa-1}(\kappa-\frac{v_\theta(\kappa-1)}{\gamma})
\tilde{b}_i(p')\leq \ddot{b}_i(p').
\end{align*}
Therefore, the claim is proved by repeatedly applying it.
\end{document}